\documentclass[twocolumn,appendixfloats,tighten]{aastex6}
\usepackage{xspace}
 
\newcommand{\lir}{L_{\rm IR}}

\newcommand{\luv}{L_{\rm UV}}

\newcommand{\msun}{~\mathrm{M_{\odot}}}
\newcommand{\lsun}{~\mathrm{L_{\odot}}}

\newcommand{\gadgettwo}{\textsc{gadget-2}\xspace}
\newcommand{\sunrise}{\textsc{sunrise}\xspace}

\newcommand{\irxbeta}{$\mathrm{IRX}-\beta$}

\newcommand{\rev}[1]{#1}
\newcommand{\revtwo}[1]{#1}

\begin{document}

\title{The IRX-$\beta$ relation:~insights from simulations}

\author{Mohammadtaher Safarzadeh\altaffilmark{1}, Christopher C. Hayward\altaffilmark{2,3,4},
and Henry C. Ferguson\altaffilmark{5}}

\shorttitle{The IRX-$\beta$ relation:~insights from simulations}
\shortauthors{Safarzadeh, Hayward, \& Ferguson}

\altaffiltext{1}{Johns Hopkins University, Department of Physics and
Astronomy, 366 Bloomberg Center, 3400 N. Charles Street, Baltimore, MD
21218, USA; \href{mailto:mts@pha.jhu.edu}{mts@pha.jhu.edu}}
\altaffiltext{2}{Center for Computational Astrophysics, Flatiron Institute, 162 Fifth Avenue, New York, NY 10010, USA; \href{mailto:chayward@simonsfoundation.org}{chayward@simonsfoundation.org}}
\altaffiltext{3}{TAPIR 350-17, California Institute of Technology, 1200 E. California Boulevard, Pasadena, CA 91125, USA}
\altaffiltext{4}{Harvard--Smithsonian Center for Astrophysics, 60 Garden Street, Cambridge, MA 02138, USA}
\altaffiltext{5}{Space Telescope Science Institute, 3700 San Martin Boulevard, Baltimore, MD 21218, USA; \href{mailto:ferguson@stsci.edu}{ferguson@stsci.edu}}

\begin{abstract}
We study the relationship between the UV continuum slope and infrared excess (IRX$\equiv L_{\rm IR}/L_{\rm FUV}$) predicted by performing dust radiative transfer on a suite of hydrodynamical simulations of galaxies. Our suite includes both isolated disk galaxies and mergers intended to be representative of galaxies at both $z \sim 0$ and $z \sim 2-3$.  Our low-redshift systems often populate a region around the the locally calibrated \citet[][M99]{M99} relation but move above the relation during merger-induced starbursts. Our high-redshift systems are blue and IR-luminous, therefore lie above the M99 relation. The value of $\beta$ strongly depends on the dust type used in the RT simulation: MW-type dust leads to significantly more negative (bluer) slopes compared with SMC-type dust. The effect on $\beta$ due to variations in the dust composition with galaxy properties or redshift is the dominant model uncertainty. The dispersion in $\beta$ is anti-correlated with SSFR and tends to be higher for the $z \sim 2-3$ simulations. In the actively star-forming $z \sim 2-3$ simulated galaxies, dust attenuation dominates the dispersion in $\beta$, whereas in the $z \sim 0$ simulations, the contributions of SFH variations and dust are similar. For low-SSFR systems at both redshifts, SFH variations dominate the dispersion. Finally, the simulated $z \sim 2-3$ isolated disks and mergers both occupy a region in the \irxbeta\ plane consistent with observed $z \sim 2-3$ dusty star-forming galaxies (DSFGs). Thus, contrary to some claims in the literature, the blue colors of high-z DSFGs do not imply that they are short-lived starbursts.
\end{abstract}


\section{Introduction} \label{sec:intro}

Dust plays a key role in many areas of astrophysics. In galaxies, it obscures emission from stars and active galactic
nuclei (AGN), thereby making it more difficult to infer the properties of these objects, such as stellar ages.
Various approaches are used to correct for the effects of dust attenuation; one popular approach is to use
the so-called `IRX-$\beta$' relation. Here, $\beta$ refers to the ultraviolet (UV) continuum slope,
which is defined by assuming that the UV regime of galaxy spectral energy distributions (SEDs) can be described
by a power law, $f_{\lambda} \propto \lambda^{\beta}$ \citep[e.g.,][]{Calzetti:1994im,M99}. Dust reddening
and older stellar populations both cause the UV continuum to be steeper and thus $\beta$ to be more positive.
IRX denotes the `infrared excess' and is defined as IRX$\equiv L_{\rm IR}/L_{\rm FUV}$, where
$L_{\rm IR}$ and $L_{\rm FUV}$ are the total infrared (IR) and far-UV luminosities, respectively.
\citet[][hereafter M99]{M99} demonstrated that local starburst galaxies exhibit a relatively tight, monotonic
relation between IRX and $\beta$. 

Recent re-calibrations of the \irxbeta\ relation prefer lower values of IRX for the same UV slope compared with
the M99 relation \citep{Overzier:2011da,Takeuchi:2012ev,Casey:2014dy,AlvarezMarquez:2016de}.  
It has been found that massive galaxies at high redshifts have similar IRX values as their low-redshift counterparts but with
bluer UV slopes \citep{AlvarezMarquez:2016de}, which implies that using the locally calibrated \irxbeta\, 
relation would cause one to underestimate the SFRs of high-mass LBGs.
Therefore, we should reconsider the similarity of the Lyman break galaxies \citep[LBGs;][]{Steidel1996}
at $z\sim2$ to local starburst galaxies based on high-redshift LBGs falling onto the M99 relation \citep{Reddy:2012ck}.

Due to the sensitivity and confusion limits \citep{de_confusion} of current
IR surveys, for many UV-selected galaxies, the IR luminosity
is unknown; thus, inferring the dust obscuration from the UV slope $\beta$
is highly desirable if one wishes to constrain the total star formation
history of the Universe \citep[e.g.,][]{Bouwens:2009ik,Bouwens:2012ht,Bouwens:2014bz,Dunlop:2012jl,Finkelstein:2012bm}.
For galaxies for which M99 relation holds, the observed slope can in principle be used to infer the
amount of UV light that has been obscured by dust, and thus one can obtain the intrinsic UV flux and total star formation
rate (SFR) from UV observations alone.\footnote{\rev{However, it is important to note that stellar populations older than
10-100 Myr can `contaminate' -- and sometimes even dominate -- both the intrinsic UV and observed IR emission of galaxies
\citep[e.g.][]{Salim:2009,Kelson:2010,Groves:2012,Leroy:2012,Smith2012,Fumagalli:2013,H14,Utomo14}. We will not
discuss this issue further in this work, but this caveat should be kept in mind.}}

If observations and/or simulations indicate that there is significant dispersion in IRX
for a fixed value of $\beta$, then it would be useful to determine whether there are other UV-optical
observable properties of a galaxy that could be used to predict its location in the \irxbeta\ plane.
It is known that (ultra-)luminous infrared galaxies [(U)LIRGs], alternatively referred to as dusty star-forming galaxies (DSFGs),
do not obey any well-defined \irxbeta\, relation
\citep{2002goldader,Bell:2002en,Howell:2010ib,Casey:2014dy}. 
Furthermore, recent studies have revealed more scatter than the original M99 work, even if (U)LIRGs are excluded.
Assuming an exponentially declining star formation history (SFH), \citet{Kong:2004uu} suggest that the dispersion in \irxbeta\,
can be explained by the ratio of the recent SFR to the SFR averaged over a longer timescale (a proxy for the SFH).
\citet{Kong:2004uu} show that the distance of galaxies from the M99 relation exhibits no correlation with the dust-sensitive $H_{\alpha}/H_{\beta}$ ratio
(i.e., the Balmer decrement), thus suggesting that dust attenuation is not responsible for the dispersion in the relation.
In contrast, \citet{Cortese:2006bg} conclude that different dust geometries can explain the observed dispersion,
but this conclusion may be affected by calibration issues \citep{Casey:2014dy}.
We discuss our results regarding the source of dispersion in Section \ref{sec:dispersion}.

The goal of this paper is to understand how galaxies might be {\it expected} to 
evolve in the \irxbeta\ plane given reasonable geometries, SFHs,
merger parameters, etc. and accounting for viewing angle-related effects.
We wish to address questions such as the following: how much of the dispersion in $\beta$ at 
fixed IRX is due to viewing angle?
How much of the dispersion is due to physical differences such as star formation histories and dust
geometries? How do galaxy mergers evolve in the \irxbeta\ plane? 
Are high-redshift DSFGs predominantly merger systems or isolated disks?

To address the above questions,
we analyze a set of 51 idealized (i.e., non-cosmological) galaxy simulations, including both mergers and isolated disks,
that are intended to be representative of both low-redshift ($z \sim 0$) and high-redshift ($z \sim 2-3$) galaxies.
In post-processing, we perform dust radiative transfer on the simulated galaxies at various times to predict
their UV--mm SEDs. This enables us to \emph{forward-model} the positions of the simulated galaxies
in the \irxbeta\, plane.
These or similar simulations have been shown to be in good agreement with the 
SEDs/colors of diverse classes of real galaxies, such as local normal star-forming galaxies \citep{Jonsson10},
(U)LIRGs \citep{Younger:2009bv,Jonsson10,Lanz14}, high-redshift DSFGs 
(\citealt{N10b,H11,H12}), obscured AGN (\citealt{Snyder:2013}) and
post-starburst galaxies \citep{Snyder11}.
We analyzed the same set of simulations in \cite{Safarzadeh:2016ew}, where we showed that the 
dispersion in the FIR SEDs of our simulated galaxies is determined primarily by the luminosity absorbed by dust
and dust mass (i.e., dust geometry is subdominant).

The remainder of this paper is organized as follows: in Section \ref{sec:simulations}, we summarize the details
of the simulated galaxy SED dataset used in this work.
Section \ref{sec:results} discusses how the SFR, observed and intrinsic UV continuum slopes, and IRX values
of some representative isolated disk and galaxy merger simulations 
evolve in the \irxbeta\, plane. Section \ref{sec:dusttype} demonstrates how the results depend on the
dust model assumed in the radiative transfer calculations. In Section \ref{sec:dispersion}, we investigate
the dispersion in the \irxbeta\, relation. Section \ref{sec:dsfgs} discusses the location of DSFGs in the \irxbeta\, plane.
In Section \ref{sec:prev_work}, we compare our results with previous work.
Section \ref{sec:implications} presents some implications for interpreting observations.
Section \ref{sec:lims} discusses the
limitations of this work and provides suggestions for future work, and Section \ref{sec:conclusions} summarizes
our conclusions.
In the appendix, we demonstrate that our results are converged with respect to the resolution of the hydrodynamical
simulations and discuss the uncertainties associated with the input SED
templates and model for sub-resolution dust structure employed in the radiative transfer calculations (both of which do
not significantly affect our results).


\section{Simulated galaxy SED dataset}\label{sec:simulations}

Our low-redshift simulation dataset was presented in \citet[][hereafter L14]{Lanz14}. The four progenitor disk galaxies span a stellar (baryonic) mass range of
$6 \times 10^8 - 4 \times 10^{10} \msun$ ($10^9 - 5 \times 10^{10} \msun$), and their properties were selected to represent typical star-forming
galaxies in the local universe \citep{Cox08}. Each of the four progenitors -- together with the 10 different possible combinations of them as merger systems --
were simulated for a single non-`special' orbital configuration for multiple gigayears (see L14 for details). The total dataset contains $\sim 6000$ SEDs. 

The details of the second set of SEDs of simulated isolated disk and merging galaxies are presented in \citet[][hereafter H13]{H13}.
For this dataset, the structural properties of the progenitor disk
galaxies were scaled to $z = 3$ following \citet{Robertson06}, with the initial gas fractions of the disks (0.6-0.8) being significantly larger
than those of the $z \sim 0$ simulations.
Because the original purpose of this suite of simulations was to model $z \sim 2-3$ DSFGs, 
the progenitor disks span a relatively narrow baryonic mass range of $\sim 1-4 \times 10^{11} \msun$, but a variety of merger orbits and mass ratios are included.
This dataset contains 37 hydrodynamical simulations, from which $\sim 46,000$ SEDs were calculated.

 The full methodology is described in the aforementioned works and references therein, so we will only summarize it here. First,
idealized isolated (i.e., non-cosmological) galaxy models were created following the method described in \citet{Springel05feedback}.
Each initial disk galaxy is composed of a dark matter halo, stellar and gaseous disks, and a supermassive black hole (SMBH); for the $z \sim 0$ simulations
only, a stellar bulge is also included. Then, the isolated galaxies and binary mergers of these galaxies were simulated using a heavily modified
version of the \gadgettwo $N$-body/smoothed-particle hydrodynamics (SPH) code \citep{Springel05gadget}.\footnote{\citet{Hayward2014arepo}
demonstrated that the results of such simulations are insensitive to the inaccuracies inherent in the traditional density-entropy formulation
of SPH, so the numerical scheme employed does not represent a significant source of error.}

The simulations include the effects of gravity, hydrodynamical interactions, and radiative heating
and cooling.\footnote{The gravitational softening lengths for
baryonic particles are $\sim 100-150$ pc, and those of the dark matter particles are $2-4$ times greater (see H13 and L14 for details).
We have confirmed that our results are converged with respect
to the resolution of the hydrodynamical simulations (Appendix \ref{S:resolution}; see also \citealt{MartinezGalarza:2016ih}).}
Star formation and stellar feedback
are incorporated via the two-phase sub-resolution interstellar medium (ISM) model of \citet{Springel03}, and BH accretion and AGN feedback are treated following
\citet{Springel05feedback}. Each gas particle is enriched with metals according to its associated SFR, assuming a yield of 0.02.
Instantaneous recycling is assumed.

The UV-mm SEDs of the simulated galaxies are forward-modeled by post-processing the outputs of the 3-D hydrodynamical
simulations at various times with the dust radiative transfer code \sunrise \citep{Jonsson06,Jonsson10}. For a given snapshot, the
\sunrise calculation proceeds as follows: the stellar and BH particles in the \gadgettwo simulation, which are the sources of radiation,
are assigned source SEDs according to their properties (age and metallicity for the star particles and luminosity for the BH particles). The metal
distribution from the simulation is projected onto an octree grid in order to calculate the dust optical depths. We use a fixed dust-to-metal
ratio of 0.4 \citep[e.g.,][]{Dwek:1998,James:2002}. The Milky Way (MW) $R_V = 3.1$ dust model of \citet{draine_li_07} is used
except for the runs with Small Magellanic Cloud (SMC)-type dust described in Section \ref{sec:dusttype}.
After the source and dust properties are specified, radiation transfer is performed using the Monte Carlo method to calculate the effects of dust absorption,
scattering, and re-emission. For each snapshot, this process yields spatially resolved UV--mm SEDs of the simulated galaxy/merger viewed
from 7 viewing angles.

\rev{Most {\sc sunrise} parameters were set to the fiducial values determined by \citet{Jonsson10}, who compared various flux ratios spanning the UV through submm
of simulated isolated disc galaxies predicted using {\sc sunrise} with those of real galaxies from the SINGS \citep{Kennicutt:2003,Dale:2007} sample, generally finding
good agreement. The SEDs of simulated galaxy mergers predicted using {\sc sunrise} with the same parameters agree well with those of local interacting galaxies
(L14) and high-redshift 24 $\mu$m-selected starbursts and AGN (Roebuck et al., submitted). The sole difference in terms of {\sc sunrise} assumptions
between the simulations used in this work and those of \citet{Jonsson10} is that in our $z = 3$ simulation suite, we set the PDR covering fraction of the \citet{Groves:2008ey}
model to 0 and employed the `multiphase off' treatment of the ISM, in which the total dust content in the simulations, rather than just that implicit in the diffuse phase
of the \citet{Springel03} sub-resolution ISM model, is considered when performing the radiative transfer. The motivation for this choice is discussed in
\citet{H11}, and we provide additional details in Appendices \ref{S:sps_model} and \ref{S:multiphase}.}

\rev{Despite employing the parameter choices motivated by the aforementioned works, to ensure that the results presented in this work are robust
(at least given the inherent limitations of the methods employed, which are discussed in
detail in Section \ref{sec:lims})}, we tested whether varying multiple potentially important assumptions in the radiative transfer calculations affected our results. Of the various parameters 
and assumptions investigated, we found that only the assumed dust composition had a significant effect on our results (i.e., the evolution in the \irxbeta\, plane);
this is discussed in detail in Section \ref{sec:dusttype}. Changing other potentially relevant assumptions typically caused the resulting $\beta$ value to differ by $\sim 0.2$ or less,
and IRX was generally negligibly affected; see Appendices \ref{S:sps_model} and \ref{S:multiphase} for details.

In Table \ref{tab:title}, we summarize the properties of the progenitors of merger systems that are studied here.

\begin{table*}
\centering
\caption{Properties of the progenitor disc galaxies.}
\begin{tabular}{llccccccc}
\hline
\hline
Name	& Redshift & $M_{\rm *,init}$				& $M_{\rm gas,init}$			& $M_{200}$				& $M_{\rm bulge,init}$			& $R_{\rm d}$ \\
		&		& ($10^{10}$ M$_{\odot}$)	& ($10^{10}$ M$_{\odot}$)	& ($10^{12}$ M$_{\odot}$)	& ($10^{10}$ M$_{\odot}$)	& (kpc)	\\
\hline
M0   & 	0	 & 	0.061 & 	0.035	&  	0.05		& 0.002	&1.12\\
M1   & 	0	 & 	0.38 & 	0.14		& 	0.20		& 0.3		&1.48\\
M2   & 	0	 & 	1.18 & 	0.33		& 	0.51		& 1.5		&1.91\\
M3   & 	0	 & 	4.22 & 	0.8		&	1.16		& 8.9		&2.85\\
b4    & 	3	 & 	0.070 & 	2.9		& 	0.86 		& 0		&2.42\\
b5    & 	3	 & 	2.0 & 	8.0		& 	2.4		& 0		&2.85\\
b5.5 & 	3	 & 	3.9 & 	16		& 	4.7		& 0		&3.7\\
b6    & 	3	 & 	7.6 & 	31		& 	9.1		& 0		&4.7\\
\hline
\end{tabular}\label{tab:summary}
\vskip 0.3cm
\label{tab:title}
\end{table*}

\section{Evolution of the simulated galaxies in the \irxbeta\, plane} \label{sec:results}
We will now discuss how the simulated galaxies evolve in the \irxbeta\, plane.
We present the results for two (one $z \sim 0$ and one $z \sim 2-3$) representative isolated disk galaxy simulations
in Section \ref{sec:isolated_disks}, and two representative merger simulations are discussed
in Section \ref{sec:mergers}.  

\begin{figure}
\centering
\vskip -0.0cm
\resizebox{3.0in}{!}{\includegraphics[angle=0]{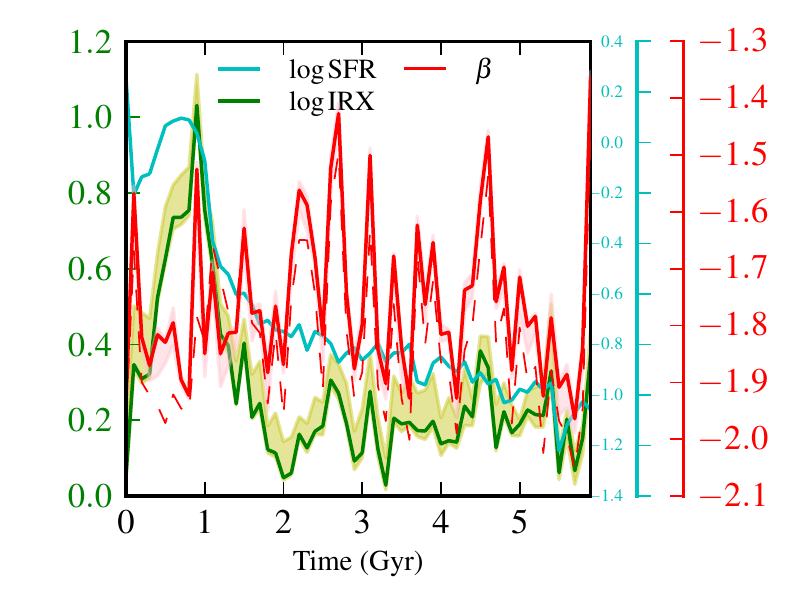}}
\resizebox{3.0in}{!}{\includegraphics[angle=0]{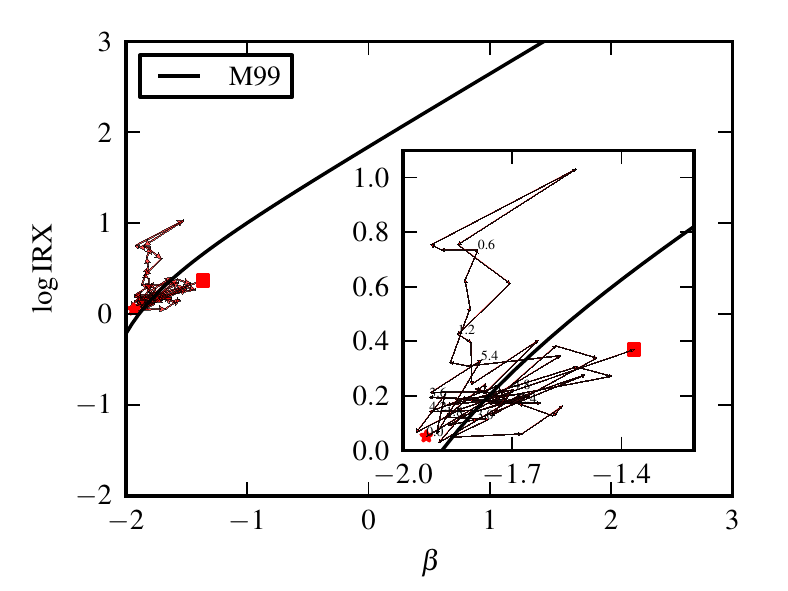}}
\caption{Results for a simulated $z \sim 0$ isolated disk galaxy, the M3 simulation of L14.
{\em Top}: The time evolution of $\log$ (SFR/M$_{\odot}$ yr$^{-1}$) (cyan), the observed and intrinsic (i.e., unattenuated) UV slopes $\beta$ (thick-solid and
thin-dashed red lines, respectively), and $\log$ IRX $\equiv \log (L_{\rm FIR}/L_{\rm UV})$ (green line). The observed $\beta$ and IRX values depend on viewing angle; the lines
correspond to the average taken over the 7 viewing angles, and the shaded pink and yellow regions show the
dispersion in the observed $\beta$ and IRX values, respectively, over the different viewing angles (see text for details).
The $\log$ IRX axis labels are provided on the lefthand side of the figure, and the $\beta$ and $\log$ SFR values are provided
on the right in red and cyan text, respectively.
{\em Bottom}: The evolution of the simulated galaxy in the \irxbeta\, plane. This figure utilizes axis ranges that will
be used for all \irxbeta\, plots in this work (see text for details); consequently, the data are concentrated in a small section of the plot.
The inset shows a zoomed-in view of this region. The solid line shows the M99 relation. The red star indicates the
initial time snapshot of the simulation, and the red square marks the final snapshot. The arrows connect individual snapshots (which are separated
by 100 Myr), and the time elapsed in Gyr is labeled for a small subset of the snapshots. At all times, this simulated disk galaxy lies
near the M99 relation; the `spur' above the relation corresponds to early in the simulation (first $\sim 1$ Gyr), when the galaxy is the most
gas-rich, highly star-forming, and dust-obscured.}
\label{fig:isolated_z0}
\end{figure}

\subsection{Isolated disks}\label{sec:isolated_disks}
Our simulation suite contains 8 isolated disk simulations, 4 of which are representative of $z \sim 0$ galaxies
and 4 of which represent $z \sim 2-3$ galaxies. The 
details of these simulations are presented in L14 and H13, respectively.

The top panel Figure \ref{fig:isolated_z0} shows the time evolution of both the attenuated
and intrinsic UV slopes, IRX and $\log$ SFR for one of the simulated $z \sim 0$ isolated disks, the M3 simulation of
L14.\footnote{We measure the observed (intrinsic) UV continuum slope from the dust-attenuated
(unattenuated) SED by least-squares fitting a functional form of $f_{\lambda}\propto \lambda^{\beta}$ to the SED
in the \rev{rest-frame} wavelength range of 1400 \AA\ $< \lambda < 2300$ \AA. We also measured the UV slope by convolving the
\rev{rest-frame} SEDs
with the GALEX FUV and NUV filter response curves, and we obtained slopes indistinguishable from those obtained by
fitting the SEDs in the \rev{rest-frame} wavelength range of $1400-2300$ \AA. The FUV luminosity is measured from the
attenuated SEDs at \rev{rest-frame} $\lambda=1600$ \AA, i.e.,
$L_{FUV}=\lambda L_{\lambda}$ at $\lambda=1600$ \AA. 
The FIR luminosity ($L_{\rm FIR}$) is defined 
by integrating the SED from 8 to 1000~$\mu$m. We then calculate $IRX\equiv L_{FIR}/L_{FUV}$.}
After an initial transient owing to the initial conditions being slightly out of equilibrium,
the SFR (the cyan line in the top panel) smoothly declines as the gas is depleted.\footnote{Because these are idealized, non-cosmological
simulations, no cosmological gas accretion is included.}
The UV slope measured from the attenuated SEDs is shown in red. It varies rapidly on a timescale of $\sim100$ Myr, 
ranging from $\sim-2$ to $\sim -1.5$, and there seems to be no long-timescale trend.
The pink shaded region shows the dispersion due to the viewing angle, which is calculated as the difference between the
84th and 16th percentiles of the distribution divided by two (for a Gaussian distribution, this is equivalent to 1$\sigma$).
For this particular simulation, the
dispersion due to viewing angle is clearly less than that due to the time evolution of the galaxy. 
The UV continuum slope measured from the intrinsic (i.e., unattenuated) SEDs
is shown with a thinner red line, which is below the thicker red line at all times.
Both of these lines are plotted on the same scale, and the values can be read from the red axis on the righthand side of the figure.
The differences between the observed and intrinsic UV slopes are modest, which suggests that dust attenuation does not
significantly alter this simulated galaxy's UV slope.
The measured IRX is indicated by the green line, and the dispersion with viewing angle is denoted by the yellow shaded region.
The IRX axis values are specified by the green numbers on the \emph{left} side of the figure.
The IRX value is high ($\sim 10$) in the early stages of the simulation, when the galaxy is still gas-rich. As the gas reservoir is
depleted and dust is locked into stars, thus decreasing the amount of dust attenuation, IRX becomes of order unity.

The bottom panel of Figure \ref{fig:isolated_z0} shows the time evolution of this simulation in the \irxbeta\, plane.
The main figure uses the axis ranges that will be used in all \irxbeta\, plots shown in this work.\footnote{The ranges were
set based on the \irxbeta\, region spanned by real galaxies \citep[e.g.,][]{Takeuchi:2012ev,Casey:2014dy}. Use
of common axes enables the reader to more easily compare the amount of evolution exhibited by the different
simulations and the magnitude of the differences that result from the choice of dust model, for example.}
The inset shows a zoomed-in view of the region traversed by the galaxy.
The initial snapshot of the simulation is marked by
the red star, and the final snapshot is indicated by the red square. The arrows connect one snapshot
to the next, tracing the galaxy in time. The thick black line is the M99 relation. 
As the simulation evolves, the galaxy wanders around the M99 relation and crosses it many times, but it
is typically close to the relation. The `spur' above the relation corresponds to the first
$\sim 1$ Gyr, when the galaxy is still gas-rich and dust-obscured, as indicated by the high IRX values shown in the
top panel.

\begin{figure}
\centering
\vskip -0.0cm
\resizebox{3.0in}{!}{\includegraphics[angle=0]{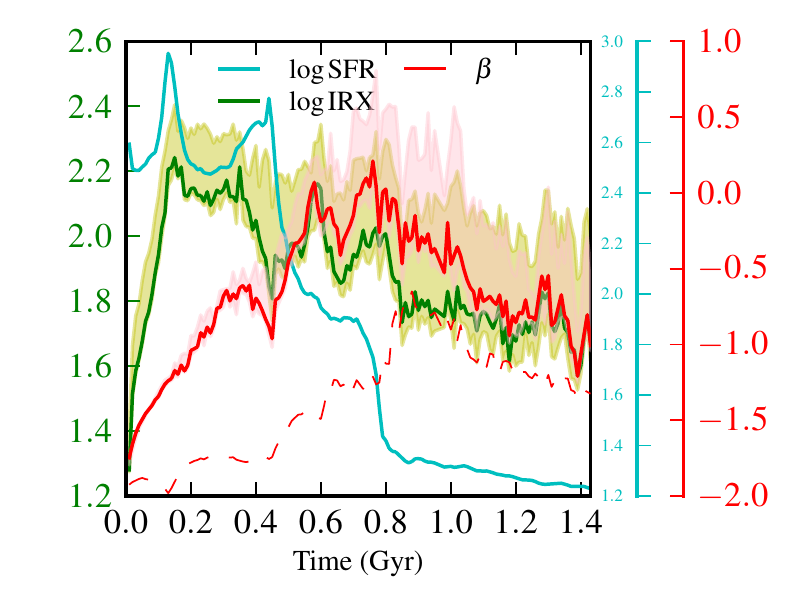}}
\resizebox{3.0in}{!}{\includegraphics[angle=0]{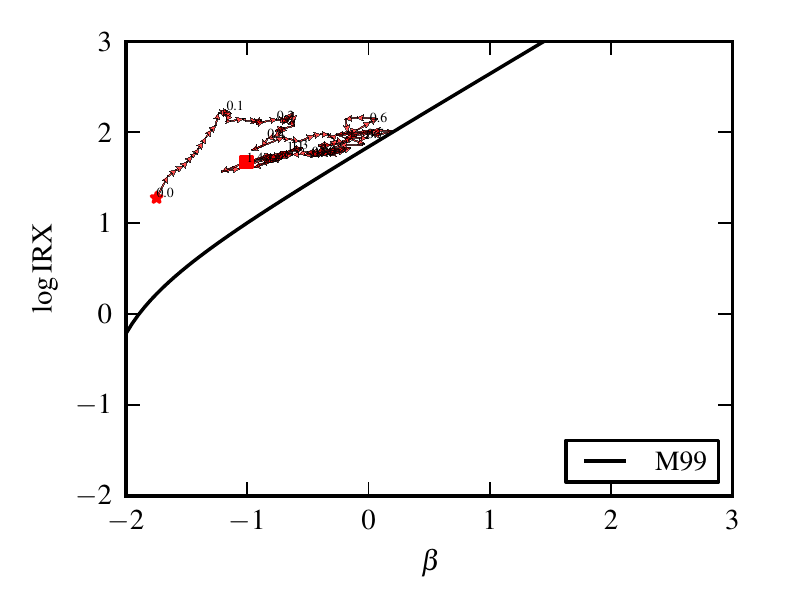}}
\caption{The time evolution (\emph{top}) and \irxbeta\, (\emph{bottom}) plots shown in Figure \ref{fig:isolated_z0} but for 
a simulated $z \sim 2-3$ isolated disk, b6 from H13.
The difference between the UV slopes of the attenuated (thick-solid red line) and unattenuated (thin-dashed red line) SEDs is more pronounced compared
with the $z \sim 0$ isolated disk shown in Figure \ref{fig:isolated_z0}. The simulated $z \sim 2-3$ isolated disk galaxy tends to lie above
the M99 relation.}
\label{fig:isolated_z3}
\end{figure}

Figure \ref{fig:isolated_z3} shows the results for one of the $z \sim 2-3$ isolated disk simulations from H13, b6.
This simulation differs from that shown in Figure \ref{fig:isolated_z0} in a few important aspects: first,
its baryonic mass is $4 \times 10^{11}$ M$_{\odot}$, whereas that of the $z \sim 0$ simulation is $5 \times 10^{10}$ M$_{\odot}$.
The $z \sim 2-3$ disk galaxy has an initial gas fraction $f_g \equiv M_{\rm gas}/(M_{\rm stars} + M_{\rm gas}) = 0.8$,
whereas the $z \sim 0$ disk has $f_g = 0.16$.
The simulated $z \sim 0$ disk galaxy contains a bulge, which acts to stabilize the gas disk, whereas the $z \sim 2-3$ disk galaxy does not.
The star particles in the initial conditions are assigned ages and metallicities appropriate for the assumed redshift; thus,
a significant fraction of the initial stellar mass of the $z \sim 2-3$ disk is comprised of young and intermediate-age stars,
whereas the stellar mass $z \sim 0$ disk is always dominated by old stellar populations.
Finally, the structural properties of this galaxy (e.g., dark matter halo and disk scalelength) have been scaled to
$z = 3$ (see H13 for details). The above differences imply that the $z \sim 2-3$ disk has a much higher gas surface density,
and thus SFR surface density, than does the $z \sim 0$ disk. 

The top panel of Figure \ref{fig:isolated_z3} shows that the time evolution of the SFR and IRX are qualitatively similar to those
of the $z \sim 0$ galaxy discussed above,
although both the SFR and IRX values are considerably greater than for the $z \sim 0$ disk owing to the much higher gas
surface density (and thus attenuation) in the $z \sim 2-3$ simulation. In contrast with the $z \sim 0$ simulation,
there is a long-term trend in $\beta$.
The value of $\beta$ tends to increase (i.e., the SED becomes redder) until
$\sim0.7$ Gyr, after which it decreases (the SED becomes bluer).
The reason for this qualitatively different behavior is that in the $z \sim 2-3$ simulations, a significant fraction of the
pre-existing stars have ages of a few hundred Myr and can thus contribute significantly to the UV emission.
As these stars age, the UV slope becomes redder.
The IRX values start a gradual decline after 700 Myr while the SED becomes bluer.
Another significant difference is that for the $z \sim 2-3$ simulation, the difference between the UV slopes
of the attenuated (thick red line) and intrinsic (\rev{red dashed} line) SEDs is considerable.
The attenuated and intrinsic $\beta$ values can differ by as much as $\sim 2$, which indicates that dust reddening
significantly affects the UV slope measured for this galaxy, unlike for the $z \sim 0$ galaxy. Relatedly, the viewing-angle-dependent
dispersion in $\beta$ and IRX (the pink and yellow shaded regions, respectively) is significant, again in contrast
with the $z \sim 0$ disk galaxy simulation discussed above.
It is worth noting that the difference between the observed and intrinsic $\beta$ values varies with time, and it becomes
less as the gas is depleted and the galaxy's ISM becomes less opaque to UV photons as dust is locked into stars.

The bottom panel of Figure \ref{fig:isolated_z3} summarizes the evolution in the \irxbeta\, plane. The galaxy tends to lie above the
M99 relation, and it comes close to the relation only at around the time when it experiences a sharp decline in SFR at $\sim0.7$ Gyr.
For such a galaxy, using the measured UV slope and the M99 relation to correct for dust would cause one to significantly underpredict
the IR luminosity, sometimes by more than an order of magnitude.


\subsection{Galaxy mergers}\label{sec:mergers}

\begin{figure}
\centering
\vskip -0.0cm
\resizebox{3.0in}{!}{\includegraphics[angle=0]{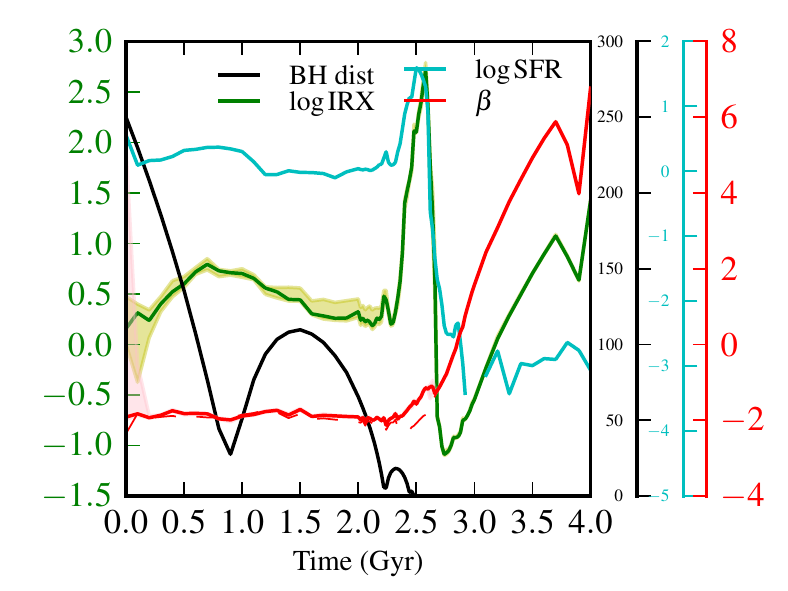}}
\resizebox{3.0in}{!}{\includegraphics[angle=0]{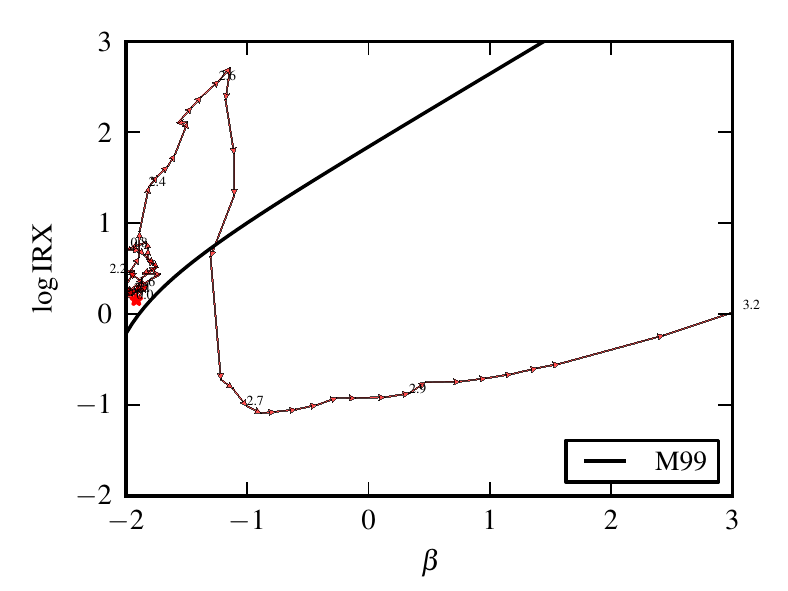}}
\caption{The time evolution (\emph{top}) and \irxbeta\, (\emph{bottom}) plots shown in Figure \ref{fig:isolated_z0} but for a simulated $z \sim 0$ equal-mass merger system,
M3M3e from L14. \emph{Top:} The black line in this panel corresponds to the separation of the central supermassive BHs in kpc; this is a proxy for merger stage.
There is a strong starburst induced near final coalescence (at $\sim 2.5$ Gyr), immediately after which
star formation is quenched because of gas consumption and AGN feedback. The time evolution of IRX is similar to that of the SFR.
$\beta$ has a relatively constant value of $\sim -2$ until the starburst, when it increases. Post-coalescence, $\beta > 0$ (i.e., the UV color
is very red) owing to the star formation being quenched. The observed and intrinsic $\beta$ values are almost identical except
for during the starburst, and the variation in IRX and $\beta$ with viewing angle is small; both of these results indicate that dust attenuation
has minor effects on the UV slope in this case despite the system being a ULIRG at coalescence. \emph{Bottom:} The merger is initially
near the M99 relation, but it moves significantly above it during the coalescence-phase starburst. Subsequently, it moves rapidly below the relation
as star formation is quenched and then moves to the right as its stellar population ages.}
\label{fig:merger_z0}
\end{figure}

\begin{figure}
\centering
\vskip -0.0cm
\resizebox{3.0in}{!}{\includegraphics[angle=0]{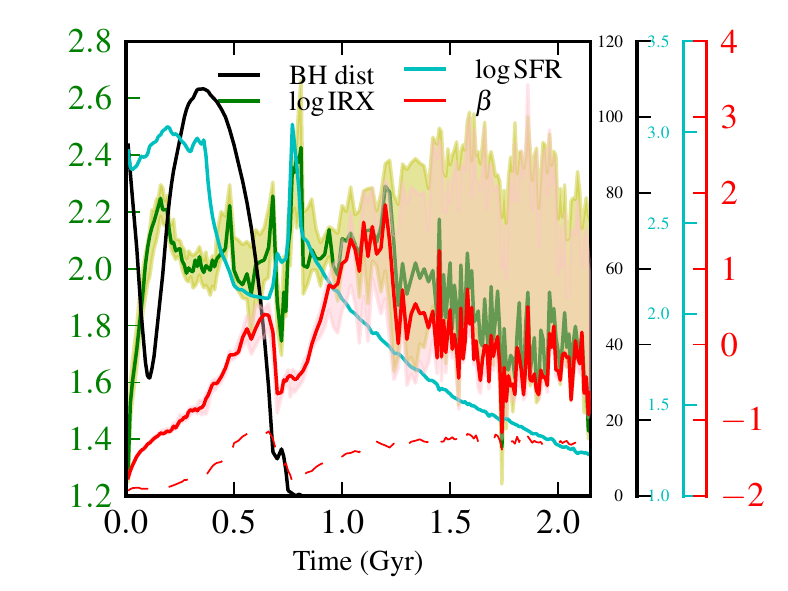}}
\resizebox{3.0in}{!}{\includegraphics[angle=0]{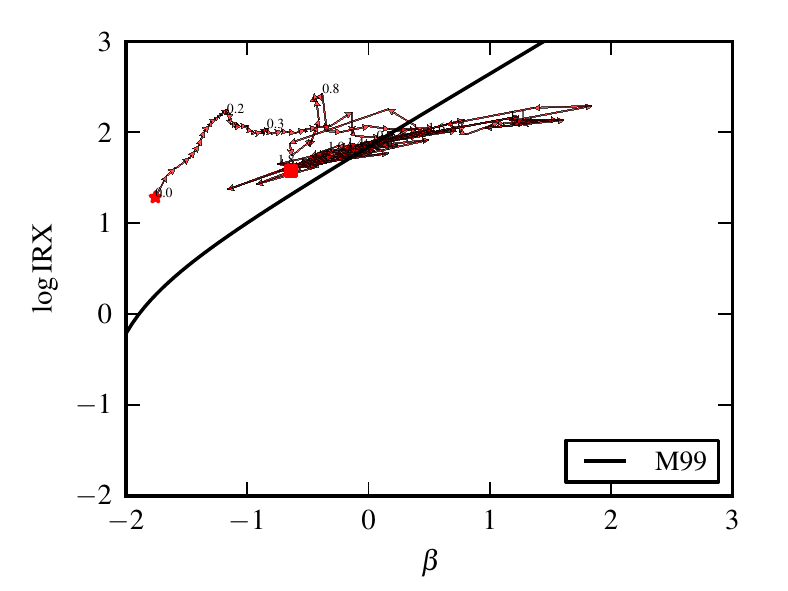}}
\caption{Similar to Figure \ref{fig:merger_z0} but for one of the simulated $z \sim 2-3$ equal-mass mergers from H13, b6b6m.
\emph{Top:} There is a strong starburst at merger coalescence ($\sim 800$ Myr), after which the SFR decreases but not as rapidly
as in the $z \sim 0$ merger presented above. At the end of the simulation, SFR $\sim 20$ M$_{\odot}$ yr$^{-1}$. As for the $z \sim 0$
merger, the evolution of IRX is similar to that of the SFR. However, the evolution of both the observed and intrinsic $\beta$ values differ
qualitatively from the $z \sim 0$ case: the intrinsic $\beta$ (\rev{red dashed} line) increases as the relatively young stars present in the initial conditions age,
decreases during the starburst, and then increases again. The time evolution of the observed $\beta$ (thick red line) is similar except after
$\sim 1.2$ Gyr, when it decreases because dust reddening is becoming less significant.
The difference between the observed and intrinsic UV slopes is large throughout the simulation, reaching
as high as $\sim 3$. This indicates that dust strongly affects the UV slope of this system. The viewing-angle-dependent dispersion in IRX and $\beta$
(yellow and pink shaded regions, respectively) is very large in the post-starburst phase. \emph{Bottom}: The system starts above
the M99 relation and moves closer to it until the coalescence-phase starburst, when it moves back above the relation. The system then
crosses the relation and moves back onto it, where it remains.}
\label{fig:merger_z3}
\end{figure}


Our simulation suite includes 10 $z \sim 0$ mergers (see L14 for details) and 32 $z \sim 2-3$ mergers (see H13 for details).
Figure \ref{fig:merger_z0} shows the time evolution of the observed and intrinsic UV slopes, IRX, $\log$ SFR and the separation of the central supermassive BHs
(a proxy for merger stage) for the M3M3e; the evolution of the other $z \sim 0$ merger simulations in which a strong starburst is induced is similar.
There is a strong increase in the SFR at $\sim2.3$ Gyr, when the two galaxies are separated by less than 10 kpc. 
The peak of the starburst occurs at $\sim2.5$ Gyr. This phase is followed by a quiescent, `red and dead' phase owing to gas consumption during the starburst and
AGN feedback. The time evolution of IRX is similar to that of the SFR. Interestingly, after $\sim 2.7$ Gyr, IRX starts to increase because the intermediate-age
and old stellar populations that dominate the dust heating in this phase \citep{H14} emit increasingly less 1600 \AA\ photons.
Both the observed and intrinsic $\beta$ values are relatively constant ($\sim -2$)
until the coalescence-phase starburst. During the starburst, the observed (intrinsic) $\beta$ increases (decreases), and both increase in the
post-starburst phase because star formation has been quenched and the stellar populations are passively evolving. The two $\beta$ values are
similar except during the starburst phase, and even then, the difference is modest compared with the $z \sim 2-3$ simulations. This indicates that
dust reddening does not significantly affect the UV slope in this system despite the system having IRX $>1$ (i.e., $L_{\rm FIR} > L_{\rm UV}$)
for much of its evolution and being a ULIRG for a brief time during the starburst. For the same reason, the dispersion in IRX and $\beta$ due to viewing angle
is small.

The time evolution of the galaxy in the \irxbeta\, plane is shown in the bottom panel of Figure \ref{fig:merger_z0}. As in the previous \irxbeta\, plots, the points corresponding
to different snapshots are connected by arrows to show the direction of time.
The initial time snapshot is indicated with a red star, and the end of the simulation is marked with a red square. The system remains close to the M99
relation up until the start of the coalescence-phase starburst, during which there is a sharp increase in IRX without a significant change in the UV continuum slope.
Consequently, the system moves above the relation. When star formation is quenched, the system moves rapidly below the relation, and it then moves to the right
as its stellar populations ages.\footnote{Note that because of our use of common axes for all \irxbeta\, plots, the system moves off the plot at
3.2 Gyr.}

The time evolution of one of the $z \sim 2-3$ equal-mass mergers (b6b6m from H13) is shown in Figure \ref{fig:merger_z3}.
The top panel shows that the starburst induced at merger coalescence occurs at $t \sim 0.8$ Gyr. In contrast with the $z \sim 0$ merger
examined above, star formation is not fully quenched: at the end of the simulation, the SFR is still $>10$ M$_{\odot}$ yr$^{-1}$.
The IRX value tends to track the SFR. For example, during the coalescence-induced starburst, IRX increases sharply, as
in the $z \sim 0$ merger shown in Figure \ref{fig:merger_z0}. Pre-coalescence, the UV continuum slope, $\beta$, increases from $\sim -1.8$ to $\sim 0.3$,
and it then decreases by $\sim 1$ during the coalescence phase. It subsequently increases from $\sim -0.7$ to $\sim 1.8$, indicating that the galaxy
transitions from blue to red, and then it decreases to $\sim -0.3$ over the remainder of the simulation. Owing to the lack of quenching, the galaxy remains
blue. The difference between the observed (thick red line) and intrinsic (\rev{red dashed} line) UV slopes is large throughout the simulation, reaching
as high as $\sim 3$. This indicates that dust strongly affects the UV slope of this system. The viewing-angle-dependent dispersion in IRX and $\beta$
(yellow and pink shaded regions, respectively) is very large in the post-starburst phase.

The bottom panel of Figure \ref{fig:merger_z3} shows the evolution of the $z \sim 2-3$ merger in the \irxbeta\, plane. Initially, the system lies above
the M99 relation. During the local minimum in the SFR at $t \sim 0.6$ Gyr, it moves onto the relation, and it subsequently moves back above
the relation during the starburst induced at final coalescence of the system.
Shortly after coalescence, it crosses the relation, and a few 100 Myr later, it returns to near the relation, where it remains until the end of the simulation
because the galaxy is still actively forming stars at a rate $>10$ M$_{\odot}$ yr$^{-1}$.


\section{Impact of dust composition}\label{sec:dusttype}

In the above analysis, the MW-type dust model of \citet{draine_li_07} was used in the radiative transfer calculations. It is worthwhile to investigate how the dust composition
affects the predicted UV slope and IRX value. In particular, the SMC extinction curve of \citet{draine_li_07} differs considerably from the MW curve
in the UV: whereas the MW extinction curve exhibits a `bump' at $\sim 2175$ \AA, the SMC curve does not. Thus, the two dust models can
lead to different dust reddening in the UV and consequently different observed UV slopes for a fixed intrinsic UV slope (i.e., fixed stellar population).
The potentially drastic impact of dust grain composition on the positions of galaxies in the \irxbeta\, plane has been discussed in the literature
\citep[e.g.,][]{Bell:2002en,Shapley:2011hy,Mao:2014bl}, but to the best of our knowledge, it has not been explored using the combination of hydrodynamical simulations and
dust radiative transfer calculations that we employ in this work.

\begin{figure}
\centering
\vskip -0.0cm
\resizebox{3.0in}{!}{\includegraphics[angle=0]{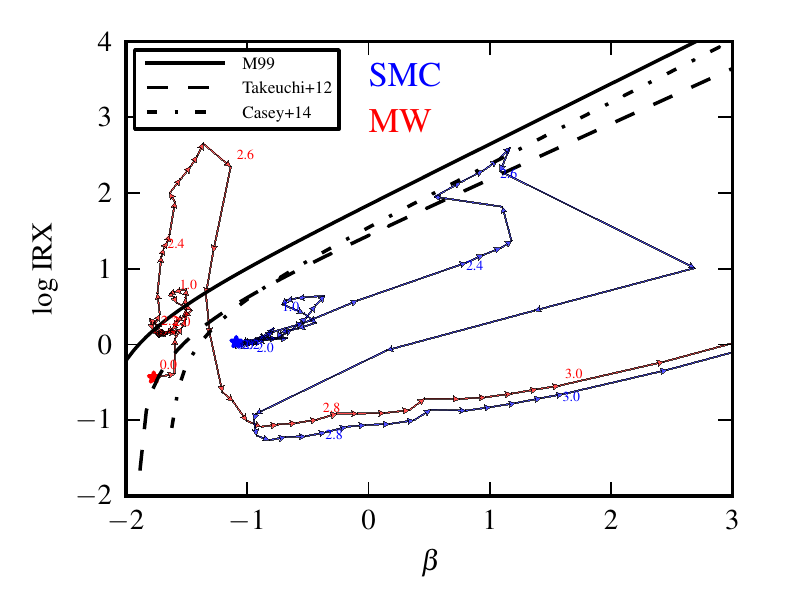}}
\resizebox{3.0in}{!}{\includegraphics[angle=0]{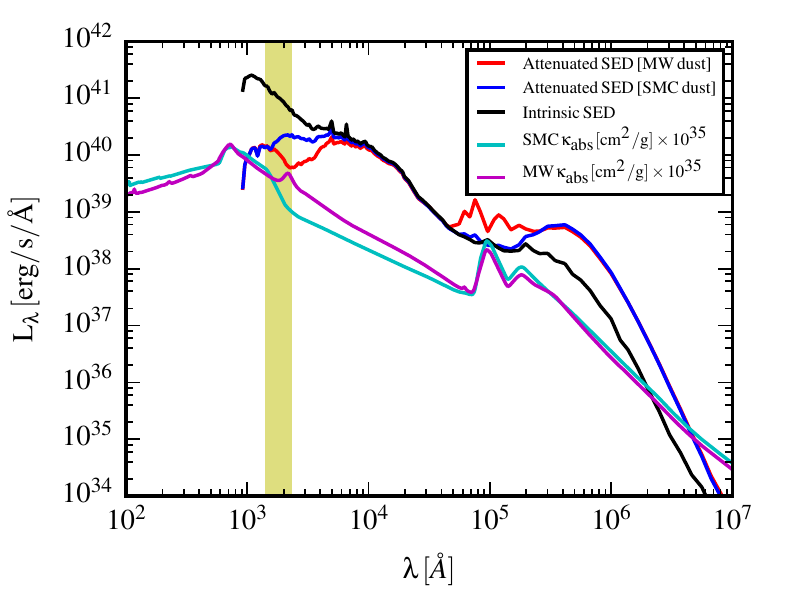}}
\caption{Demonstration of the impact of the assumed dust grain model on the final IRX and $\beta$ values.
{\emph Top}:~The evolution of the M3M3e $z \sim 0$ major merger simulation in the \irxbeta\, plane when 
MW-type (red arrows) and SMC-type (blue arrows) dust is used in the radiative transfer calculations.
For the first $\sim 2.7$ Gyr of the simulation, when the
system is actively forming stars, the assumed dust model \emph{qualitatively} changes the evolution
of the merger in the \irxbeta\, plane. $\beta$ is greater (i.e., the SED is redder) when SMC-type dust is assumed;
IRX is relatively insensitive to the assumed dust model except for low IRX values.
 {\emph Bottom}:~Comparison of the \rev{rest-frame} SEDs of the M3M3e simulation at $t=2.4$ Gyr.
The intrinsic SED of the galaxy is shown in black. The attenuated SEDs (including dust re-emission)
predicted when MW-type and SMC-type dust are assumed are shown in red and blue, respectively.
The yellow region indicates the wavelength range that is used to compute the UV slope.  
The absorption cross-section per unit mass of the MW (magenta line) and SMC (cyan line) dust models (renormalized as indicated in the legend)
are also shown. It is clear that the two SEDs have very different UV slopes: $\beta$ is negative (positive)
when MW (SMC) dust is assumed. The extinction curves clearly show that the difference between the UV slopes obtained with the two dust models
is due to the presence of the UV bump at 2175 \AA\ in the MW curve and the change in the slope of the SMC curve
in this wavelength range.}
\label{fig:dust_model}
\end{figure}

To investigate the impact of the dust grain composition, we performed radiative transfer on the $z \sim 0$ major merger simulation discussed above,
M3M3e from L14, assuming SMC-type dust instead of the fiducial MW-type dust. The results of these radiative transfer calculations are compared with
those of the standard MW-dust run in Figure \ref{fig:dust_model}. The top panel shows how the assumed dust grain model impacts 
the evolution of the merger in the \irxbeta\, plane. The blue arrows follow the evolution of the system 
when SMC-type dust is assumed, whereas the red arrows follow the evolution of the same system modeled with MW-type dust
(this track is identical to that shown in the bottom panel of Figure \ref{fig:merger_z0}). For the first $\sim 2.7$ Gyr of the simulation, when the
system is actively forming stars (see Figure \ref{fig:merger_z0}), the assumed dust model has a very significant impact on the position of the
system in the \irxbeta\, plane. In general, $\beta$ is greater (i.e., the SED is redder) when SMC-type dust is assumed. For this reason, at
fixed time, the merger can lie above the M99 relation if MW-type dust is assumed in the radiative transfer calculation but below it if SMC-type dust
is used. The effect is greatest during the coalescence-phase starburst ($t \sim 2.4 - 2.6$ Gyr): when MW dust is assumed, the merger-driven starburst
causes the system to move above the relation, but when SMC dust is assumed, the system moves \emph{closer} to the relation during the
starburst.

The bottom panel of Figure \ref{fig:dust_model} illustrates why the assumed dust model has such a drastic impact on the system's position in
the \irxbeta\, plane. The black line is the intrinsic SED of the system at $t = 2.4$ Gyr, and the red (blue) line is the attenuated SED obtained
when MW-type (SMC-type) dust is used in the radiative transfer calculation. 
The yellow shaded region denotes the wavelength range that is used to compute the UV slope (1300 \AA\ $<\lambda<2300$ \AA).
The magenta and cyan curves show the absorption cross-section per unit mass (in arbitrary units) for the MW- and SMC-type dust
models, respectively.
As is evident from comparing the SEDs, the 2175 \AA, feature present in the MW curve results in a more negative UV slope compared with 
the SMC dust model. The SMC curve starts to decrease more steeply with increasing wavelength near where the 2175 \AA\ feature becomes evident
in the MW curve, which exacerbates the difference in the UV slopes of the two SEDs. The difference between the two observed UV slopes,
which is completely due to the assumed dust composition, is approximately 2. Although there is a 
large difference in the UV slope due to the dust grain model, the effect on IRX is modest for the following reasons:
when IRX is high, the vast majority of UV photons from the young stars that dominate the intrinsic UV luminosity are absorbed regardless of whether MW- or SMC-type
dust is assumed. Thus, $L_{\rm FIR}$ is insensitive to the dust model.
Because the observed UV is dominated by unobscured stars (i.e., the UV and IR emission are essentially decoupled), $L_{\rm UV}$
is also not significantly affected. Thus, when IRX is high, it is not affected by the dust model.
For example, comparing the $t=2.4$ Gyr points in the \irxbeta\, plane in the top panel reveals that the two SEDs have the same IRX value but
very different UV slopes. Thus, the difference in $\beta$ is the primary driver of the difference in the \irxbeta\ evolution.

After the starburst, when star formation has been quenched, the choice of dust model matters little. The MW track tends to be slightly above
the SMC track because for fixed dust column density, the MW dust model yields a greater optical depth in the UV--optical regime. When IRX is low (as
it is in this phase), and thus the galaxy is not opaque to UV photons, the increase in the optical depth caused by assuming MW-type rather
than SMC-type dust can lead to a non-negligible increase in the IR luminosity and decrease in the observed luminosity, thus causing a non-negligible
(but small) increase in IRX.
 
\rev{Note that as shown in the bottom panel of Figure \ref{fig:dust_model}, the MW and SMC dust extinction curves are almost identical over the
wavelength range of $\sim 500-1500$ \AA. Consequently, for a given dust column density, the UV slope in this wavelength range should
be reddened by the same amount regardless of whether the dust is MW-like or SMC-like. For this reason, it may be possible to fit the UV slope
between the Lyman limit and 1500 \AA rather than in the canonical range (1300 \AA\ $<\lambda<2300$ \AA) to mitigate the uncertainty associated
with dust composition.}

\section{Dispersion in the UV slope}\label{sec:dispersion}

\begin{figure}
\centering
\vskip -0.0cm
\resizebox{3.0in}{!}{\includegraphics[angle=0]{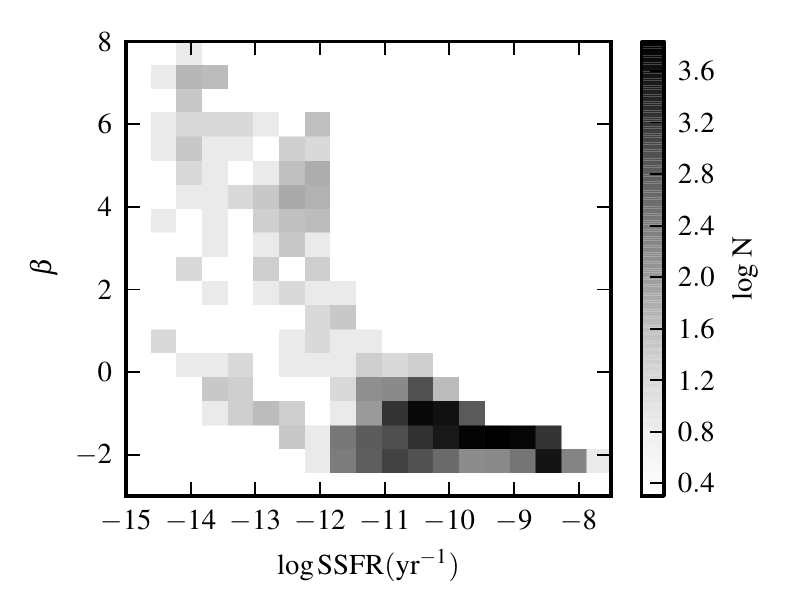}}
\resizebox{3.0in}{!}{\includegraphics[angle=0]{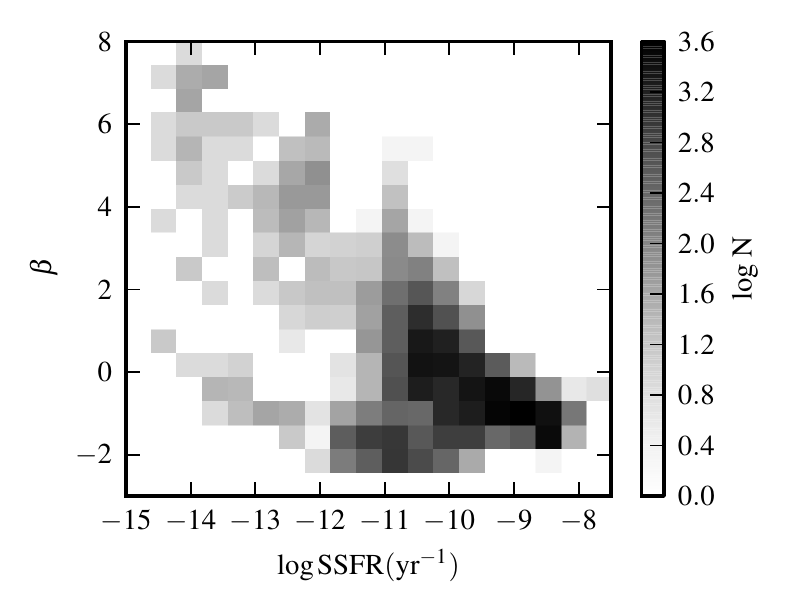}}
\caption{2D histograms of the intrinsic (\emph{top}) and observed (i.e., attenuated; \emph{bottom}) $\beta$ values versus
$\log$ (SSFR/yr$^{-1}$) for all simulated galaxies analyzed in this work (i.e., both the $z \sim 0$ and $z \sim 2-3$ datasets).
Note that the colorbar indicates the logarithm of the number of SEDs in a given pixel.
Both the intrinsic and observed $\beta$ values are anti-correlated with SSFR because in more actively star-forming galaxies,
young, massive stars dominate the luminosity and cause the UV slope to be blue. The scatter at fixed SSFR is also anticorrelated
with SSFR because the past SFH does not affect the UV slope if very young stars dominate the UV luminosity.
A comparison of the two histograms reveals that for galaxies with high SSFR, dust attenuation causes the observed UV slope to be redder than the intrinsic slope.
For passive systems, $\beta$ is essentially unaffected by dust attenuation. The bifurcation evident for SSFR $\la 10^{-11}$ yr$^{-1}$
is due to the inclusion of both low- and high-redshift simulations: at fixed SSFR, the $z \sim 0$ simulations tend to have lower $\beta$ values.}
\label{fig:dispersion}
\end{figure}

\begin{figure}
\centering
\vskip -0.0cm
\resizebox{3.0in}{!}{\includegraphics[angle=0]{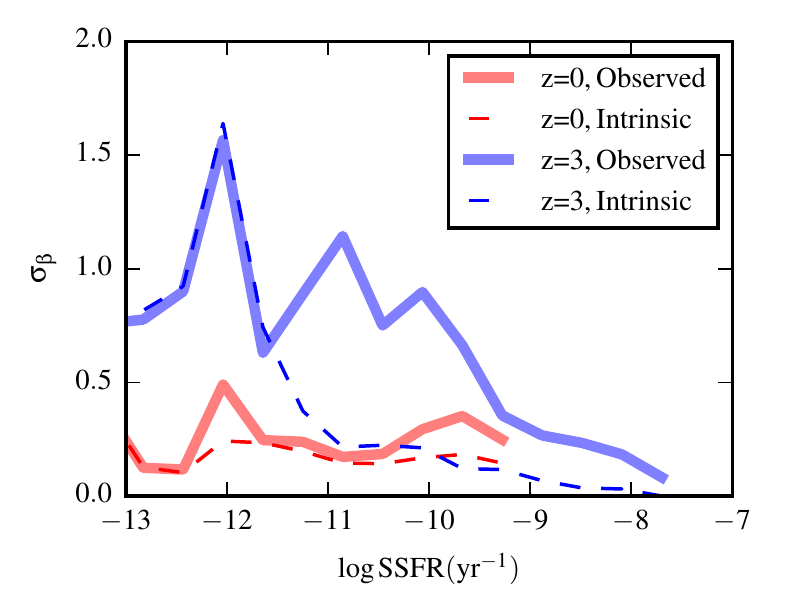}}
\caption{Dispersion in $\beta$ versus $\log$ (SSFR/yr$^{-1}$). $\sigma_{\beta}$ is defined by dividing the difference between
the 16th and 84th percentile values of the $\beta$ distribution (including all simulations and viewing angles) in a given SSFR bin by 2.
The red (blue) lines correspond to the $z \sim 0$ ($z \sim 2-3$) simulations. The solid (dashed) lines denote the dispersion in the
observed (intrinsic) $\beta$ values. The dispersion in both the intrinsic and observed $\beta$ values is anticorrelated with SSFR,
as already observed from Figure \ref{fig:dispersion}. In the $z \sim 0$ simulations with high SSFR, SFH variations and dust
attenuation contribute comparably to the dispersion; in lower-SSFR systems, SFH history variations dominate. In contrast, in the
$z \sim 2-3$ simulations with SSFR $\ga 10^{-11.5}$ yr$^{-1}$, the $\beta$ variation due to dust attenuation
(related to both viewing angle-related effects and differences in dust geometry among different time snapshots and simulations) is much greater
than that due to SFH variations. For actively
star-forming galaxies, the dispersion in the observed $\beta$ is greater for the $z \sim 2-3$ simulations than for the $z \sim 0$ simulations.
Our results suggest that even if low-redshift galaxies obey a relatively tight \irxbeta\, relation, this should not be the case
at higher redshifts.}
\label{fig:dispersion_summary}
\end{figure}

Various factors can in principle lead to dispersion in the \irxbeta\, plane, including dust geometry \citep{Cortese:2006bg},
SFH \citep{Kong:2004uu}, the mean age of the stellar population \citep{Mao:2012fi,Grasha:2013hd} or stellar mass 
\citep{AlvarezMarquez:2016de}. \citet{Boquien:2012dl} studied a set of face-on normal star forming spiral galaxies 
to understand why normal star-forming galaxies deviate from starburst galaxies in terms of their positions in the \irxbeta\, plane.
They found that the intrinsic dispersion in the UV slope correlates most strongly with the distance
from the M99 relation. \citet{Kong:2004uu} suggest that the dispersion results from differences in the
SFHs of galaxies (by considering the birth rate parameter $b$, which is the ratio of the current to past SFR;
note that this parameter can be quite sensitive to how one defines `current' when the SFR is highly variable; e.g., \citealt{2015arXiv151003869S}). 
Both studies point to the importance of underlying SFH for the observed dispersion of galaxies in the \irxbeta\, plane.
\citet{AlvarezMarquez:2016de} studied a stacked sample of LBGs at $z\sim3$ and concluded that
stellar mass best predicts deviations from the \irxbeta\, relation. They found that more-massive LBGs have bluer colors and higher dust attenuation.
This is in agreement with the trend observed by \citet{Casey:2014dy} for DSFGs.
\citet{Mao:2012fi} studied spatially resolved normal star-forming galaxies from the SINGS \citep{Kennicutt:2003,Dale:2007} sample by separating a galaxy into
UV-emitting clusters and background disk and bulge regions. The found the mean stellar population age
contributes significantly to the dispersion in the \irxbeta\, plane, and the location of the galaxy as a whole in the \irxbeta\, plane
is determined largely by the local background regions.

We can distinguish between scatter due to different SFHs and that due to dust geometry by comparing
the dispersions in the intrinsic and observed (i.e., attenuated) $\beta$ values because the intrinsic UV slope depends
only on the SFH, whereas the latter also includes the viewing angle-dependent effects of dust.
In Figure \ref{fig:dispersion}, we show 2D histograms of the intrinsic (\emph{top}) and observed (\emph{bottom})
$\beta$ values versus specific SFR (SSFR$\equiv$SFR/M$_{\star}$). All simulations analyzed in this work
(i.e., both isolated disks and mergers at both $z \sim 0$ and $z \sim 2-3$) are shown here. Both the intrinsic and observed $\beta$
values are anticorrelated with SSFR. This implies that in the most actively star-forming systems, young, massive
stars dominate the UV emission and thus make the UV slope blue. Moreover, the dispersion at fixed SSFR is
anticorrelated with SSFR, which implies that the past SFH affects $\beta$ less in actively star-forming
systems. A comparison of the two panels reveals that
for galaxies with SSFR $\ga 10^{-12}$ yr$^{-1}$, dust attenuation causes $\beta$ to increase (i.e., the UV slope
to become redder), whereas for systems with lower SSFR values, the intrinsic and observed $\beta$ distributions are
similar. The bifurcation evident for SSFR $\la 10^{-11}$ yr$^{-1}$ is due to the inclusion of both low- and high-redshift
simulations: at fixed SSFR, the $z \sim 0$ simulations tend to have lower $\beta$ values.

Figure \ref{fig:dispersion_summary} shows the dispersion in $\beta$, $\sigma_{\beta}$, versus SSFR. $\sigma_{\beta}$
is computed by binning the SEDs in terms of their corresponding SSFR values and then dividing the difference
between the 16th and 84th percentiles by two; for a normal distribution, this measure would equal the standard
deviation. $\sigma_{\beta}$ thus incorporates variations due to viewing angle, the time evolution of a given system, and
the differences among the simulations. The solid (dashed) lines correspond to the observed (intrinsic) $\beta$ values, and red
(blue) denotes $z \sim 0$ ($z \sim 2-3$). This figure reinforces our conclusion from Figure \ref{fig:dispersion}
that the dispersions in both the intrinsic and observed $\beta$ values are anticorrelated with SSFR. Considering
the $z \sim 0$ simulations, we note that the dispersions in the intrinsic and observed $\beta$ values differ by a modest
amount, and the difference is greater at high SSFR.
Computing the difference between the intrinsic and observed $\sigma_{\beta}$ values in quadrature, which
characterizes the contribution of dust attenuation to the dispersion (related to both viewing angle-related effects and
differences in dust geometry among different time snapshots and simulations), we find that dust adds at most 0.4 to $\sigma_{\beta}$,
and it typically adds $\la 0.2$. Thus, in the actively star-forming $z \sim 0$ simulations, dust attenuation and
SFH variations contribute comparably to the dispersion in the observed $\beta$ values, whereas in systems with lower SSFR,
SFH variations dominate the dispersion.

In contrast, for the $z \sim 2-3$ simulations, the difference between the intrinsic and observed $\beta$ dispersions
is significant. Again computing the difference in the $\sigma_{\beta}$
values in quadrature, we find that the variation due to dust attenuation dominates that due to SFH variations
in all bins with SSFR $\ga 10^{-11.5}$ yr$^{-1}$, in contrast with the $z \sim 0$ simulations.
For systems with lower SSFR values, the intrinsic and observed $\beta$ dispersions are almost
identical, which indicates that SFH variations dominate the dispersion. Finally, we note that for actively
star-forming galaxies, the dispersion in the observed $\beta$ values at fixed SSFR is greater for the $z \sim 2-3$
simulated galaxies than for those intended to be representative of $z \sim 0$ galaxies. Consequently, our results
suggest that even if low-redshift galaxies obey a tight \irxbeta\, relation (i.e., the M99 relation), this is unlikely to
be the case at higher redshift. Moreover, the above results suggest that the sought-after `second parameter'
(i.e., what property best predicts deviations from the M99 relation) may depend on the galaxy type and redshift
considered.

\section {\revtwo{Relationship between IRX and UV optical depth}} \label{sec:tau_uv}

\begin{figure}
\centering
\vskip -0.0cm
\resizebox{3.75in}{!}{\includegraphics[angle=0]{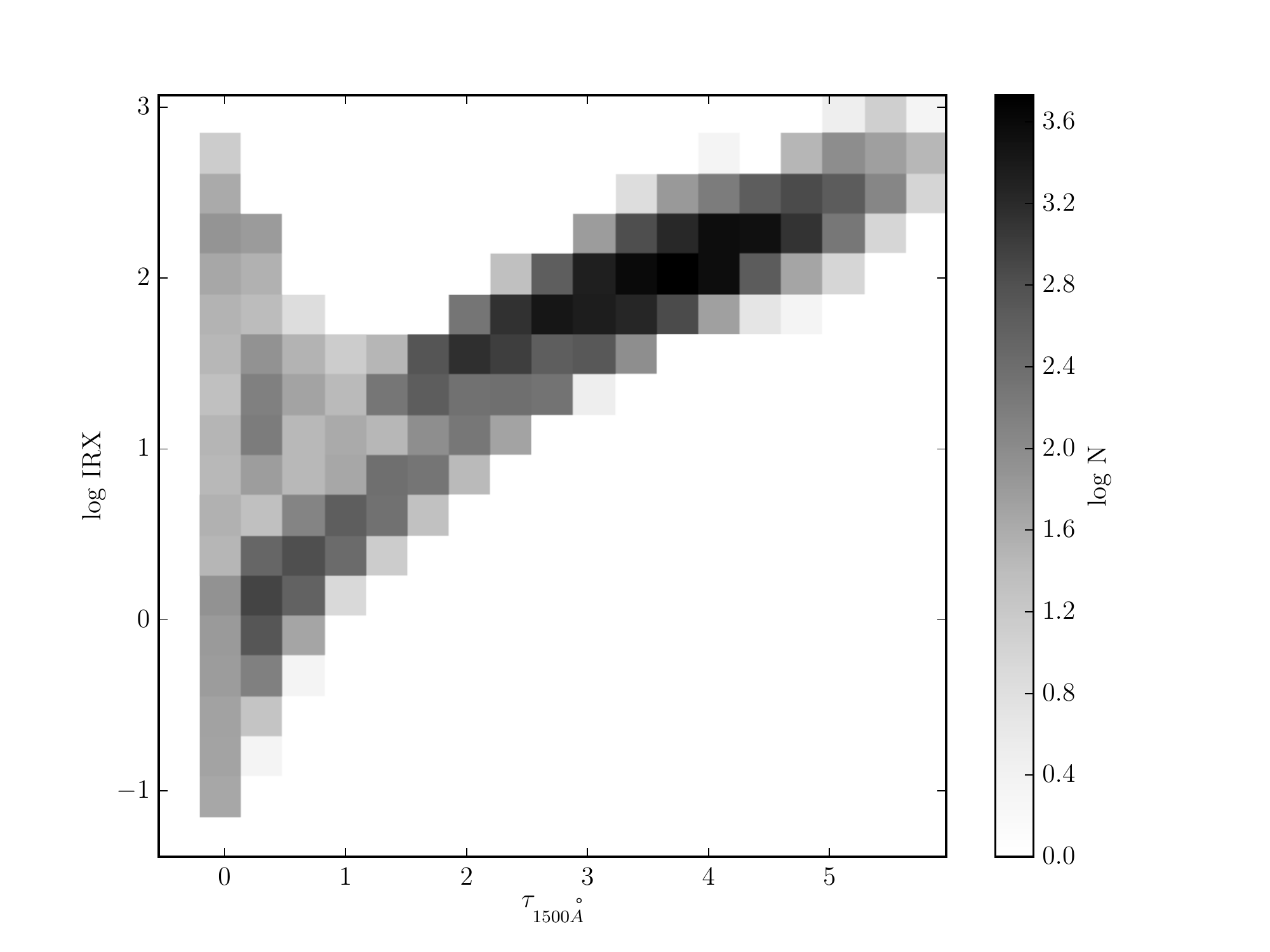}}
\caption{\revtwo{2-D histogram of the logarithm of IRX versus effective optical depth at 1500 \AA, $\tau_{\rm 1500 \AA}$. The colorbar indicates the number of simulation
snapshots in each bin. For $\tau_{\rm 1500 \AA} \ga 1$, the correlation between the two quantities is relatively tight. However, for fixed
$\tau_{\rm 1500 \AA} \la 1$, the regime in which the UV continuum escape fraction is non-negligible, IRX can vary by multiple orders of magnitude.
This large scatter implies that the UV continuum escape fraction cannot be reliably inferred from the value of IRX.}}
\label{fig:irx_vs_tau}
\end{figure}

\revtwo{One benefit of our simulations is that because the intrinsic spectrum is perfectly known, we can directly
measure the effective optical depth at arbitrary wavelengths. For this work, it is of interest to consider how the
optical depth in the UV is related to IRX.
We specifically consider the optical depth at 1500 \AA,
$\tau_{\rm 1500 \AA} \equiv -\ln(L_{\rm 1500,observed}/L_{\rm 1500,intrinsic})$.
Figure \ref{fig:irx_vs_tau} shows the distribution of the simulated galaxies in the $\log {\rm IRX}-\tau_{\rm 1500 \AA}$
plane. The bins are color-coded according to the logarithm of the number of galaxies in the bin, as indicated by the
colorbar. For $\tau_{\rm 1500 \AA} \ga 2$ (i.e., when escape fraction at 1500 \AA\ is $\la 10$ percent), $\log$ IRX
and $\tau_{\rm 1500 \AA}$ are tightly correlated. However, for fixed $\tau_{\rm 1500 \AA} \la 2$, IRX can vary by
multiple orders of magnitude. Consequently, our simulations indicate IRX cannot be used to reliably infer the UV
optical depth in the regime of most interest (when the UV escape fraction is high).
}

\section {DSFGs in the IRX-$\beta$ plane} \label{sec:dsfgs}

Many DSFGs at $z \sim 2-3$ have very blue UV continuum slopes (i.e., negative $\beta$ values) and high IRX values \citep{Casey:2014dy},
which makes them lie above the M99 relation.
Very blue and dusty systems were also found in a study of 16 individually \textit{Herschel} PACS 100-$\mu m$- and 160-$\mu m$-detected 
LBGs at $z\sim3$ \citep{Oteo:2013ey}. Although these samples are biased by the IR selection, the results may indicate
that massive galaxies tend to have bluer UV slopes and higher IRX values compared with less-massive galaxies \citep{AlvarezMarquez:2016de}.
Moreover, where DSFGs reside in the \irxbeta\, plane may yield insights into the origin of their extremely high SFRs:~\citet{Casey:2014dy} argue that the
fact that DSFGs lie above the M99 relation is evidence that they are short-timescale starbursts rather than `main sequence' galaxies (i.e., galaxies
near the approximately linear SFR--stellar mass relation).

\begin{figure}
\centering
\vskip -0.0cm
\resizebox{3.0in}{!}{\includegraphics[angle=0]{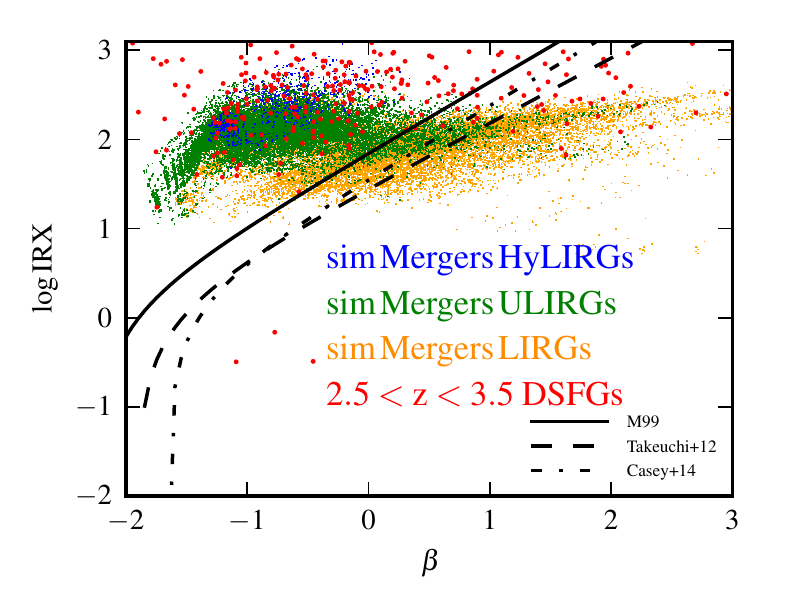}}
\resizebox{3.0in}{!}{\includegraphics[angle=0]{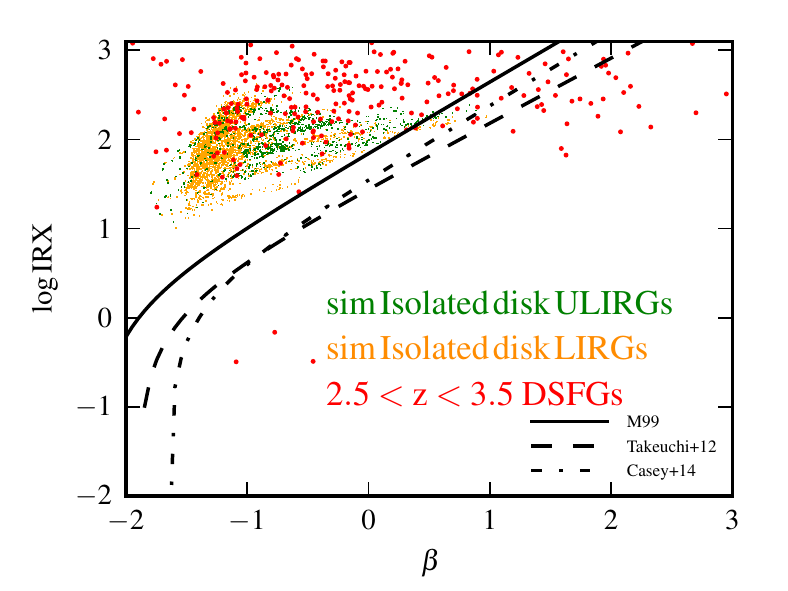}}
\caption{The positions of our simulated $z \sim 2-3$ DSFGs in the \irxbeta\, plane. The mergers (isolated disks) are shown in the \emph{top}
(\emph{bottom}) panel. The points are color-coded according to IR luminosity: orange, green, and blue points correspond to simulated LIRGs
($10^{11} < \lir/\lsun < 10^{12}$), ULIRGs ($10^{12} < \lir/\lsun < 10^{13}$) and hyper-LIRGs (HyLIRGs; $\lir > 10^{13} \lsun$), respectively.
In both panels, the red points correspond to the observed $z \sim 2.5-3.5$ DSFGs from \citet{Casey:2014dy}.
It is evident that both the simulated mergers and isolated disk can become (U)LIRGs during their evolution. However, only the mergers can reach 
$\lir > 10^{13} \lsun$ and $\log$ IRX $>2.5$. Because the simulated isolated disks overlap with observed DSFGs in the \irxbeta\, plane,
the fact that some real DSFGs lie well above the M99 relation is not evidence that they are powered by short-lived starbursts.}
\label{fig:dsfgs}
\end{figure}

We can directly address this question using our simulation suite.
In Figure \ref{fig:dsfgs}, we show the locations of our simulated $z \sim 2-3$ DSFGs in the \irxbeta\, plane; the top panel shows the results of the
simulated $z \sim 2-3$ mergers, whereas the bottom panel shows the simulated $z \sim 2-3$ isolated disk galaxies.
The orange, green, and blue points correspond to simulated LIRGs ($10^{11} < \lir/\lsun < 10^{12}$), ULIRGs ($10^{12} < \lir/\lsun < 10^{13}$)
and hyper-LIRGs (HyLIRGs; $\lir > 10^{13} \lsun$), respectively. (There are no blue points in the bottom panel because the simulated isolated
disks never attain $\lir > 10^{13} \lsun$.)
In both panels, the red points correspond to real $z \sim 2.5-3.5$ DSFGs selected from the
COSMOS survey \citep{Scoville2007cosmos}, the IRX and $\beta$ values of which were presented in \citet{Casey:2014dy}.

In both panels, the regions in the \irxbeta\, plane spanned by the real and simulated DSFGs are broadly consistent, although for $\log$ IRX
$\gtrsim 2.5$, the specific simulations used here do not span the full range of $\beta$ spanned by real DSFGs, which can be both slightly
bluer and significantly redder than the simulated galaxies \citep[see also][]{Wuyts2009}. For the mergers (\emph{top}),
the offset from the M99 relation tends to increase with IR luminosity; all of the simulated HyLIRGs are well above the M99 relation,
which indicates that despite having effectively all of their luminosity absorbed and reradiated by dust ($\lir/\luv \gtrsim 100$), they have
blue UV continuum slopes. The reason for our simulated DSFGs having blue UV colors is that in these IR-luminosity-selected systems, the UV
and IR are essentially decoupled: the
IR luminosity is powered by stars that are invisible in the UV, whereas the UV emission originates from a `frosting' of unobscured young stars.
This holds for both the simulated mergers and isolated disks.

Notably, the bottom panel shows that our simulated gas-rich $z \sim 2-3$ isolated disk galaxies are similar to DSFGs in terms of their location in
the \irxbeta\, plane. These galaxies are \emph{not} undergoing merger-driven starbursts by construction. Thus, our results demonstrate that
the position of DSFGs in the \irxbeta\, plane does not necessarily imply that they are short-lived starbursts, contrary to the claim of
\citet{Casey:2014dy}; instead, DSFGs well above the M99 relation can be steadily star-forming, massive, gas-rich disks as long as they are
sufficiently dust-obscured, which is a natural outcome for high-gas-fraction, metal-enriched galaxies. Consequently, high-z DSFGs
may represent a heterogeneous population of massive, gas-rich, steadily star-forming galaxies, merger-induced starbursts, and obscured
AGN \citep[e.g.,][]{Hopkins2010,H11,H12,H13,HB13,2015DaCunha,Koprowski2016}. We do find that all of the
simulated galaxies with $\lir > 10^{13} \lsun$ (see also \citealt{Hopkins2010}) and, of the lower-luminosity simulated galaxies,
those with the highest IRX values ($\log$ IRX $\gtrsim 2.5$) are exclusively mergers, and in the simulations, these HyLIRGs are typically powered
by combination of merger-induced starbursts and obscured AGN (Roebuck et al., submitted).
However, we cannot determine whether these are physical limits
or simply artifacts of the specific parameter space spanned by the simulations.

However, it should be noted that the IRX and $\beta$ values plotted in Figure~\ref{fig:dsfgs} are those obtained when MW-type dust is used
in the radiative transfer calculations. As we have shown in section \ref{sec:dusttype}, using MW-type dust yields more negative UV slopes 
than when SMC-type dust is assumed. Consequently, if we were to assume SMC-type dust when performing the radiative transfer calculations, our
simulated galaxies might not populate the low-$\beta$, high-IRX region that is occupied by a significant fraction of real DSFGs.
Thus, it is possible that the location of DSFGs in the \irxbeta\, plane may encode information regarding their dust composition. This is an interesting
topic that could be explored with future simulations.

\section {comparison with previous work} \label{sec:prev_work}

We now compare our results with those of some previous theoretical works.
The evolutionary tracks of galaxies in the \irxbeta\, plane were previously studied 
in the context of semi-analytic models (SAMs) coupled with radiative transfer calculations assuming simplified geometries \citep[GRASIL;][]{1998Silva} by \citet[][hereafter G00]{Granato:2000}.
G00 find that in the early stages of a starburst, the UV slopes of their model galaxies become bluer (more negative).
When the SFR decreases, and consequently older stellar populations dominate the intrinsic UV slope,
the system evolves towards more positive (redder) $\beta$ and lower IRX values.
In G00's model, the strength of the burst defines the evolutionary path on the \irxbeta\, plane.
Stronger starbursts have only slightly negative or positive (red) UV slopes ($\beta \gtrsim -0.5$), initially because they experience high attenuation
on timescales longer than the exponential decay timescale of the SFR and later because the stellar population has turned old. Their
IRX values are generally $\gtrsim 10$. In contrast, weaker starbursts span a larger 
dynamical range in the \irxbeta\, plane because these galaxies quickly exhaust their gas, which causes the attenuation to decrease rapidly
and makes them evolve towards low IRX and negative $\beta$ values.
We note that G00 assume MW dust, as have we in our fiducial calculations, and can reproduce the Calzetti attenuation law, despite the
conclusion of \citet{Gordon:1997jd} that SMC-type dust is needed to reproduce the Calzetti law. 

Unlike G00, when a strong burst occurs in our simulations (e.g., the M3M3e simulation shown in Figure~\ref{fig:merger_z0}), the system moves almost
vertically in the \irxbeta\, plane because IRX increases by a large amount, whereas the UV slope becomes only slightly redder. The two results might seem
contradictory, but the difference lies in how we define the onset of starburst. If consider the evolution of the M3M3e merger system 
starting with the peak of the starburst (during the coalescence phase), which corresponds to G00's definition (because they assume an
exponentially declining SFH for the burst), we see a similar trend: the galaxy maintains a relatively flat UV slope while moving toward
lower IRX values. The further evolution of the system when the stellar population becomes old is similar to the starbursts modeled in G00.
The starbursts modeled by G00 lie on the M99 relation as long as the age of the starburst is $\lesssim 50$ Myr. However, a comparison of
figure 7 of G00 with Figure \ref{fig:dsfgs} indicates that the starbursts in the G00 model never occupy the high-IRX, low-$\beta$
region spanned by both our simulated DSFGs and observed DSFGs. This discrepancy may be a result of the parameter space that
they considered, their assumption of an exponentially declining SFH for bursts, or/and the simplified geometry assumed
in their radiative transfer calculations.

\citet{KhakhalevaLi:2016uw} performed radiative transfer on galaxies formed in cosmological simulations from the
``Cosmic Reionization on Computers'' (CROC) project \citep{Gnedin2014} to compare the UV and IR properties
of their simulated $z > 5$ galaxies, including their positions in the \irxbeta\, plane, with observations.
They find that in order to match the observed UV luminosity function at $z>5$, they must include the effects
of dust destruction via sublimation in supernova shocks; otherwise, their galaxies are too dusty and thus UV-faint
to explain the observed UV luminosity function. Their simulated galaxies lie near the M99 relation, although
IRX can vary by approximately an order of magnitude at fixed $\beta$. The locus of their simulated galaxies
in the \irxbeta\, plane is consistent with some $z > 5$ galaxies for which both $\beta$ and IRX have been
measured \citep{Capak2015}, but the galaxies with $\log$ IRX $\gtrsim 0.5$ from \citet{Capak2015} and the
$z \sim 7.5$ galaxy presented by \citet{Watson2015} lie outside the region spanned by the simulated galaxies.
Because our simulations and theirs cover disjoint regions of parameter space in terms of redshift and mass, it
is not possible to make detailed comparisons between our results and theirs. However, their results highlight
the potential importance of dust destruction for accurately predicting the UV properties of simulated galaxies
and provide motivation for future studies that include more complex treatments of dust production, growth and destruction.

\section{Implications for observations} \label{sec:implications}

Our simulation results suggest that there is not a tight relation between IRX and $\beta$ that applies to all
galaxies at all redshifts. The $z \sim 0$ simulated galaxies tend to lie near the M99 relation, which was
determined based on observations of galaxies in the local Universe. However, during merger-induced
starbursts, even the $z \sim 0$ galaxies can depart significantly from the relation: because IRX increases
significantly but $\beta$ is almost unaffected, the $z \sim 0$ merger-induced starbursts, such as that shown in
Figure \ref{fig:merger_z0}, tend to lie above the M99 relation. The $z \sim 2-3$ simulations -- of both disk galaxies
and mergers -- deviate more significantly from the M99 relation: even when they are not undergoing starbursts,
they tend to lie above the relation.

These results are consistent with observational works that have demonstrated
that some classes of galaxies deviate significantly from the M99 relation. As already discussed above, observed DSFGs tend
to lie above the M99 relation and do not exhibit any relationship between IRX and $\beta$
\citep[e.g.,][]{2002goldader,Bell:2002en,Howell:2010ib,Casey:2014dy}. Conversely, some galaxies lie
significantly below the M99 relation (i.e., have lower IRX than expected from the relation given their $\beta$).
This was demonstrated for the lensed galaxies cB58 at $z\sim2.7$ \citep{Pettini:1999dy} and the Cosmic Eye
\citep{Smail:2007bq} at $z\sim3.07$. Similarly, \citet{Reddy:2012ck} found that young systems
(age $<100$ Myr) at $z\sim2$ lie below the M99 relation.

Moreover, we have demonstrated that the dust composition -- specifically the strength of the 2175 \AA\ feature --
can have a very dramatic effect on the UV slope.
Consequently, extrapolation of the locally calibrated \irxbeta\, relation to regimes for which it is not observationally
constrained may be problematic if, e.g., the dust composition of high-redshift galaxies is significantly different
from that of $z \sim 0$ galaxies. How the strength of the UV bump varies with galaxy properties and redshift is still poorly understood,
but there are some useful observational constraints. The commonly employed \citet{Calzetti:1994im} attenuation law, which was
defined based on observations of local-Universe starburst galaxies, does not exhibit a UV bump.
Based on a study of $\sim$10,000 $z<0.1$ star-forming galaxies, \citet{Battisti:2016jb} 
derived an average attenuation curve that does \emph{not} have a significant 2175 \AA\ feature.
However, \citet{Motta:2002dx} presented evidence for the UV bump in a
gravitationally lensed normal early type galaxy at $z=0.83$. Since then, other detections for high-redshift galaxies
have been reported in the literature \citep{Noll:2009cw,Buat:2011dz,Kriek:2013kx}.
\citet{Conroy:2010it} studied attenuation as a function of inclination using a sample of nearby disk galaxies. Trends due to 
dust attenuation alone can be identified in such studies because stellar population properties are unrelated to inclination.
The trends observed by \citet{Conroy:2010it} suggest the presence of the 2175 \AA\ feature with an amplitude
$\sim$80\% of the canonical MW value.
Studying $\sim30$ galaxies at $z>1$, \citet{Buat:2011dz} detected a significant UV bump at 2175 \AA\ but with
an amplitude $\sim35\%$ of that of the MW extinction curve. For a sample of galaxies in the redshift range $0.5<z<2$,
\citet{Kriek:2013kx} found evidence for a UV bump whose strength increases with the slope of the attenuation curve.
Given the above mixed and perhaps even contradictory results, more effort clearly needs to be invested in
understanding how the composition of dust varies across cosmic time and among different galaxy populations.

\revtwo{Because the intrinsic spectrum is well-described by a single power law over the rest-frame wavelength range
$\sim1100-3600$ \AA, in the absence of dust attenuation, one can in principle recover the intrinsic $\beta$ equally well using
any rest-frame wavelengths within the aforementioned range (but choosing more widely separated wavelengths
is of course beneficial for providing a larger `lever arm').
It may be possible to minimize or eliminate the effect of the 2175 \AA\ feature identified in Section \ref{sec:dusttype}
by carefully selecting the
bands employed to calculate $\beta$ according to a galaxy's redshift.
Specifically, for some redshifts, one could `straddle'
the feature using e.g. the NUV-$u$ color: at $z = 0.4$, this would be equivalent to fitting a power law to the fluxes at
1570 and 2600 \AA, thus minimizing the influence of the 2175 \AA\ feature. At $z = 1$, these bands correspond
to rest-frame wavelengths of 1100 and 1820 \AA. For higher redshifts, other filter combinations could be used.
Although the difference in the slope of the attenuation curve between the MW and SMC models would not be
addressed by such an approach, much of the uncertainty owing to the dust composition, which is especially
poorly constrained for high-redshift galaxies, would be ameliorated.}

The \irxbeta\, relation is often applied to classes of galaxies (in terms of properties such as mass and redshift) for which the
\irxbeta\, relation is not directly constrained. The above discussion suggests that it is unlikely that
there is a universal \irxbeta\, relation; instead, variations in both global galaxy properties and dust composition
can cause significant deviations from the M99 relation. Constraints on the SFH of the
Universe at $z \gtrsim 3$ are almost exclusively based on dust-correcting the UV luminosity using the M99 relation to obtain
the SFR \citep[e.g.][]{Bouwens:2009ik,Bouwens:2012ht,Dunlop:2012jl,Finkelstein:2012bm,Bouwens:2014bz}.
Use of the M99 relation can cause one to underestimate the SFRs of galaxies that lie above the M99 relation,
such as DSFGs, possibly by multiple orders of magnitude. If such galaxies contribute non-negligibly to the SFH
of the Universe, the UV-based constraints would not recover the true SFR density. For this reason, it is crucial to obtain
direct constraints on obscured star formation via rest-frame IR observations of galaxies at higher redshift ($z \gtrsim 3$) and
with less-extreme SFRs ($\lesssim 100$ M$_{\odot}$ yr$^{-1}$) than has been possible with surveys performed with, e.g.,
the JCMT and \emph{Herschel}. ALMA is an excellent tool with which to address this challenge, and some very interesting
constraints have already been obtained \citep[e.g.,][]{Capak2015,Watson2015}, but more work needs to be done.

\section{limitations and future work} \label{sec:lims}

Although our simulations have yielded insights into the evolution of galaxies in the \irxbeta\, plane, they are
of course subject to some limitations. First, they are idealized non-cosmological simulations. The advantage of such
simulations is that they enable one to explore a significant region of the relevant parameter space while achieving the
resolution necessary to perform radiative transfer on the simulated galaxies. However, their non-cosmological nature
implies that effects such as cosmological gas accretion and subsequent mergers are not included, and the demographics
of our simulation suite are not cosmologically representative. \rev{Unfortunately, given the resolution limitations of state-of-the-art
cosmological hydrodynamical simulations (see the discussions in \citealt{Sparre2015illustris} and \citealt{Sparre2016}, for example),
such simulations are currently not useful for modeling the propagation of UV light on galaxy scales and thus the \irxbeta\, relation.}

Moreover, we employ the \citet{Springel03}
sub-resolution model for the structure of the ISM and the effects of supernova feedback. Although this model is still
widely used, simulations that resolve the small-scale structure of the ISM and feature more sophisticated, \rev{explicit} treatments of
stellar feedback have been presented (e.g., \citealt{Hopkins2014}; \citealt*{Agertz2015}). \rev{In such simulations,
properties such as disc scaleheights, which may be determined by stellar feedback-driven turbulent pressure support
\citep*[e.g.][]{FG13,HH15} or cosmic rays \citep[e.g.][]{Booth2013},
are likely more realistic than in the simulations used in this work. Altering the vertical structure of the gas -- and thus dust --
discs in the simulations could affect our predicted \irxbeta\, relation.

Additionally, given that such simulations resolve the ISM on scales
of approximately an order of magnitude smaller than our simulations, they may more faithfully capture how UV light
propagates in real galaxies than the present simulations do. \revtwo{The present simulations may not correctly capture
the relative amount of UV emission attenuated by dust local to star-forming regions instead of `diffuse' dust. The discrepancy
between the 24-to-70-\micron\ flux ratios of the simulated galaxies presented in \citet{Jonsson10} and
observed galaxies may be evidence for such a discrepancy (although the 24-\micron\ emission, in particular,
is sensitive to other uncertain aspects of the radiative transfer calculations, such as non-equilibrium emission from stochastically
heated very small grains). It will be very interesting to investigate how the balance between local and diffuse attenuation
differs in the aforementioned simulations that resolve the small-scale ISM and to compare their colors with those of real galaxies.}

Higher-resolution simulations with explicit stellar feedback
also self-consistently generate galactic winds \citep{Muratov2015}. The lack of winds in our simulations may cause e.g. the
time evolution of the gas and metal content of our galaxies to be somewhat unrealistic. However, this is less of a concern
for the idealised non-cosmological simulations used here than for cosmological simulations because the cosmological
evolution of the baryon and metal content of galaxies -- which is determined by a combination of inflows, outflows, and
self-enrichment -- is intentionally not modelled in our work. Instead, we constructed initial conditions by hand such that
our simulated galaxies have properties such as gas fractions and metallicities that are roughly consistent with
empirical constraints (see e.g. \citealt{Cox08} and \citealt{Rocha2008} for details).}

\rev{For the above reasons, it would be of great interest to repeat
our analysis using such simulations, and this work is underway. However, because of the high computational expense
of cosmological zoom-in simulations that resolve GMC scales and include explicit stellar feedback, it is a challenge
to construct run large samples of simulations for statistical studies. Moreover, the cosmological nature of the simulations
and chaotic behavior caused by explicit stellar feedback imply that controlled studies of the effect of a single parameter
(e.g. gas fraction) are challenging. For these reasons, idealised simulation suites such as that used in this work are still useful
for some studies.}

Finally, we have not employed a detailed model for dust production and destruction
but rather assumed that the dust-to-metal density ratio is constant. As \citet{KhakhalevaLi:2016uw} have noted,
the details of dust production and destruction may have important implications for the UV and IR properties of
simulated galaxies, \rev{especially in low-metallicity galaxies at high redshift}. It would thus be useful to perform an analysis similar
to ours using simulations that include
a detailed treatment of the relevant dust production and destruction channels \citep[e.g.,][]{McKinnon2016,McKinnon2016b}.

\section{conclusions} \label{sec:conclusions}

We have analyzed a set of 51 hydrodynamical simulations of idealized (i.e., non-cosmological) galaxies, including both isolated
disk galaxies and mergers at both $z \sim 0$ and $z \sim 2-3$, on which dust radiative transfer was performed in post-processing to
yield UV--mm SEDs, from which we measured the UV continuum slope ($\beta$) and IRX $\equiv \lir/\luv$. This method enables us to
forward-model the evolution of the simulated galaxies in the \irxbeta\, plane. Our primary conclusions are as follows:
\begin{itemize}
\item The simulated $z \sim 0$ isolated disk galaxies tend to reside near the M99 relation, whereas the $z \sim 2-3$ disks
are generally above it (i.e., at fixed $\beta$, they have higher IRX values than expected from the M99 relation).
\item In the simulated mergers, when a strong starburst is induced near coalescence, the systems tend to move almost vertically
in the \irxbeta\, plane because IRX increases significantly but $\beta$ remains similar. In the post-starburst phase,
if star formation is quenched via gas consumption and AGN feedback, IRX decreases rapidly, and $\beta$ increases (i.e., the UV
slope becomes redder) as the stellar population ages.
\item The dust type that is assumed in the radiative transfer calculations drastically impacts the resulting `observed' UV continuum slope. 
MW-type, for which the extinction curve exhibits a `bump' at 2175 \AA, results in a significantly more negative (i.e., bluer) UV slope
than SMC-type dust, for which the extinction curve does not exhibit the 2175 \AA\ bump and has a steeper slope in the UV compared
with MW-type dust. When the simulated galaxies are actively forming stars, the effects of the assumed dust model on the
galaxies' positions in the \irxbeta\, plane are more significant than the dispersion due to differences in galaxy
properties, such as mass and SFH, the dispersion due to viewing angle, and model uncertainties other than the dust composition.
\item The dispersion in both the intrinsic and observed $\beta$ values is anticorrelated with SSFR. Considering actively star-forming simulated galaxies,
dust attenuation dominates the dispersion in $\beta$ for the $z \sim 2-3$ simulations, whereas in the $z \sim 0$
simulations, the contributions from SFH variations and dust attenuation are similar. At low SSFR, SFH variations dominate
the $\beta$ dispersion at both redshifts. These results suggest that the sought-after `second parameter' (i.e., what property best predicts
deviations from the M99 relation) should depend on the galaxy type and redshift considered.
\item When SSFR $\ga 10^{-11.5}$ yr$^{-1}$, the dispersion in the observed $\beta$ at fixed SSFR is higher in the $z \sim 2-3$ simulations
than in the $z \sim 0$ simulations. This result indicates that
even if a relatively tight \irxbeta\, relation is obeyed by low-redshift galaxies, this should not be the case for galaxies at higher redshift.
\item \revtwo{IRX is well correlated with the effective optical depth at 1500 \AA\ when the latter quantity is $\ga 2$. However, when the UV
escape fraction is non-negligible (the regime of most interest for e.g. studies of reionization), IRX can vary by multiple orders of
magnitude at fixed optical depth.}
\item The simulated $z \sim 2-3$ DSFGs tend to lie above the M99 relation (i.e., have high IRX values
yet blue UV slopes), similar to observed $z \sim 2-3$ DSFGs (although this result may be sensitive to the assumed dust model).
Consequently, in contrast with some previous claims in the literature, the position of DSFGs in the \irxbeta\, plane is not evidence that they are powered by short-lived starburst
events; rather, they may also be `main sequence' galaxies as long as they are sufficiently dust-obscured. However, the simulations
with $\lir > 10^{13}$ or/and $\log$ IRX $\gtrsim 2.5$ are exclusively merger-driven starbursts, sometimes with a significant contribution
from obscured AGN.
\end{itemize}

Our work adds to the growing consensus that a well-defined \irxbeta\, relation is \emph{not} obeyed by all galaxy populations.
DSFGs, in particular, can deviate significantly, and inferring the SFR by dust-correcting the UV luminosities of such galaxies
using the M99 or a different \irxbeta\, relation would cause their SFRs to be underestimated, potentially by multiple orders of
magnitude. Consequently, the UV-inferred star formation rate density may be more unreliable than is generally believed.
This highlights an urgent need for current and future observatories, such as ALMA, the Large Millimeter Telescope, and the
Chajnantor Submillimeter Survey Telescope \citep{Padin2014}, to directly constrain
the obscured SFRs of galaxies at higher redshift and lower SFRs than are probed by existing surveys.

\acknowledgements

We thank Patrik Jonsson for insightful discussions. 
We are thankful to Caitlin Casey for providing us with data in electronic form. The Flatiron Institute is supported by the Simons Foundation.
CCH is grateful to the Gordon and Betty Moore Foundation for financial support.
This work was partially supported by NASA's Astrophysics Data Analysis Program under grant NNX15AE54G.
The computations in this paper were run on the Odyssey cluster supported by the FAS Division of Science, Research Computing Group at Harvard University.
This research has made use of NASA's Astrophysics Data System Bibliographic Services and the \href{www.arxiv.org}{arXiv.org} preprint server.


\bibliographystyle{aasjournal}
\bibliography{uv_evolution}

\appendix

\section{Convergence with respect to hydrodynamical resolution} \label{S:resolution}
\begin{figure}
\centering
\vskip -0.0cm
\resizebox{3.0in}{!}{\includegraphics[angle=0]{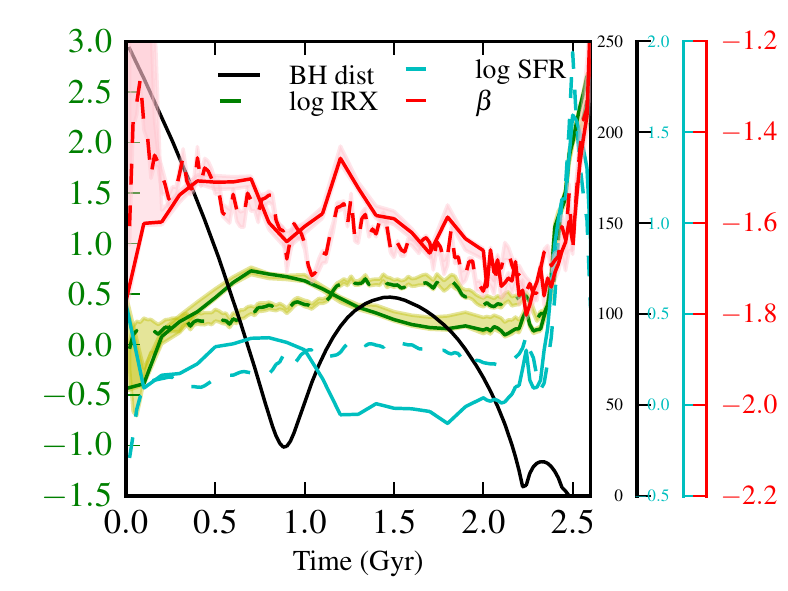}}
\resizebox{3.0in}{!}{\includegraphics[angle=0]{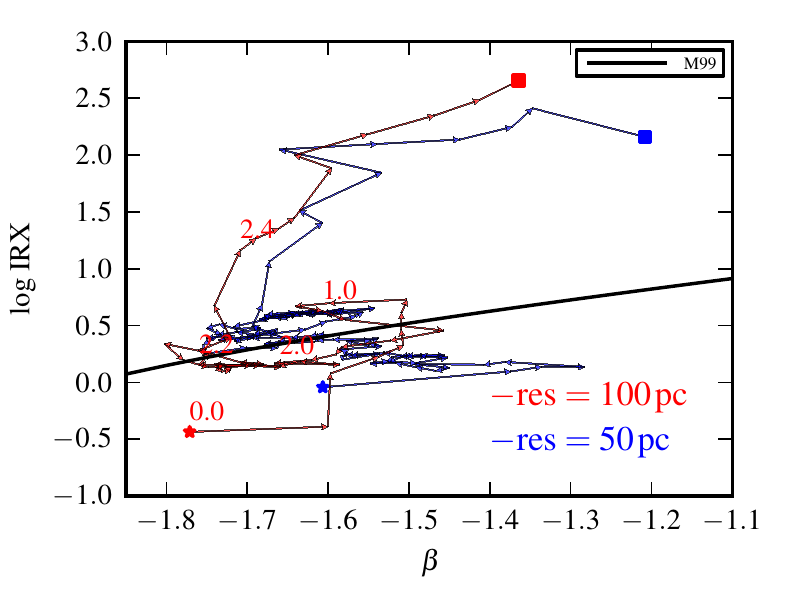}}
\caption{Comparison of the standard M3M3e $z \sim 0$ equal-mass merger with a run with 8 times the number of particles and 2 times smaller
softening lengths. \emph{Left:} time evolution of $\log$ IRX, $\beta$, $\log$ SFR, and supermassive BH separation for both the standard-resolution M3M3e simulation
(solid lines) and higher-resolution version (dashed lines). The minor differences in the SFHs of the two simulations result in minor
differences in the time evolution of IRX and $\beta$, but the qualitative evolution is the same.
\emph{Right:} evolution of the two simulations in the \irxbeta\, plane. Note that the region of the \irxbeta\, plane shown here is much smaller than that shown
in the \irxbeta\, plots in the main text because we wish to highlight the (minor) differences between the simulations. The paths of the two simulations differ
in detail but are qualitatively identical, and the differences owing to resolution are much less than the uncertainty due to the dust composition.}
\label{fig:sim_res}
\end{figure}

To confirm that our results are converged with respect to the resolution of the hydrodynamical simulations, we simulated one of the merger 
simulations (M3M3e) with twice the spatial resolution and 8 times more particles with respect to the standard M3M3e simulation
and then performed radiative transfer on the results.
Because of both computational constraints and our primary focus being actively star-forming galaxies, we ran this simulation only until
slightly after the merger-induced starburst.
Figure \ref{fig:sim_res} compares the results of this higher-resolution simulation with those of the standard M3M3e run from L14. The left panel
shows the time evolution of the SFR, observed $\beta$, and IRX; the solid (dashed) lines correspond to the standard (higher-resolution) M3M3e run.
The SFHs of the two simulations differ slightly; consequently, the evolution of $\beta$ and IRX differ somewhat. However,
the qualitative behavior is identical, and the differences are small. For example, at fixed time, the $\beta$ values of the two simulations differ by at most 0.3, and the
difference is typically $<0.1$. The right panel of Figure \ref{fig:sim_res} compares the evolution of the two simulations in the \irxbeta\, plane;
note that the region of the \irxbeta\, plane shown here is much smaller than that shown in the \irxbeta\, plots in the main text because we wish to highlight the
(minor) differences between the simulations. Although the detailed evolution of the two simulations differs due to the different SFHs,
the tracks of the two simulations in the \irxbeta\, plane are qualitatively indistinguishable, and the differences owing to resolution are much less than the uncertainty
due to the dust composition (see Section \ref{sec:dusttype}).

\section{Effects of assumptions regarding obscuration of young stars} \label{S:sps_model}

\begin{figure}
\centering
\vskip -0.0cm
\resizebox{3.0in}{!}{\includegraphics[angle=0]{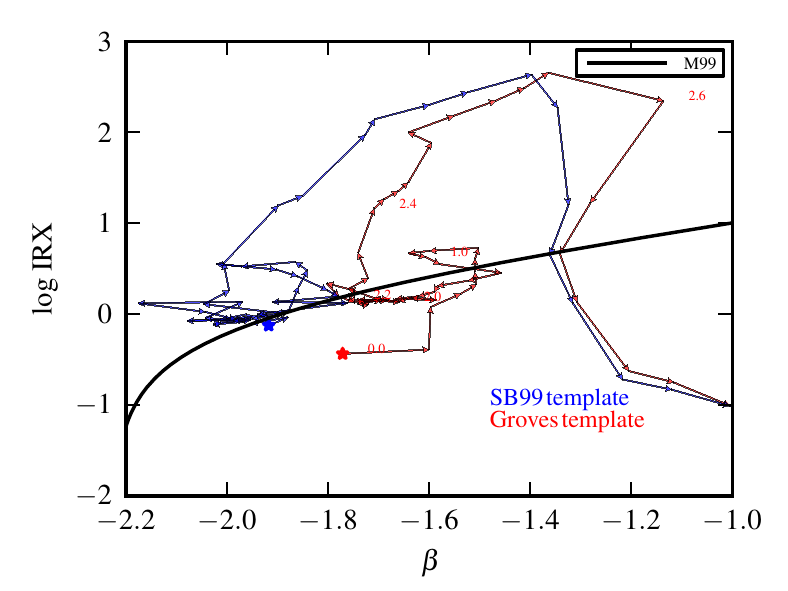}}
\resizebox{3.0in}{!}{\includegraphics[angle=0]{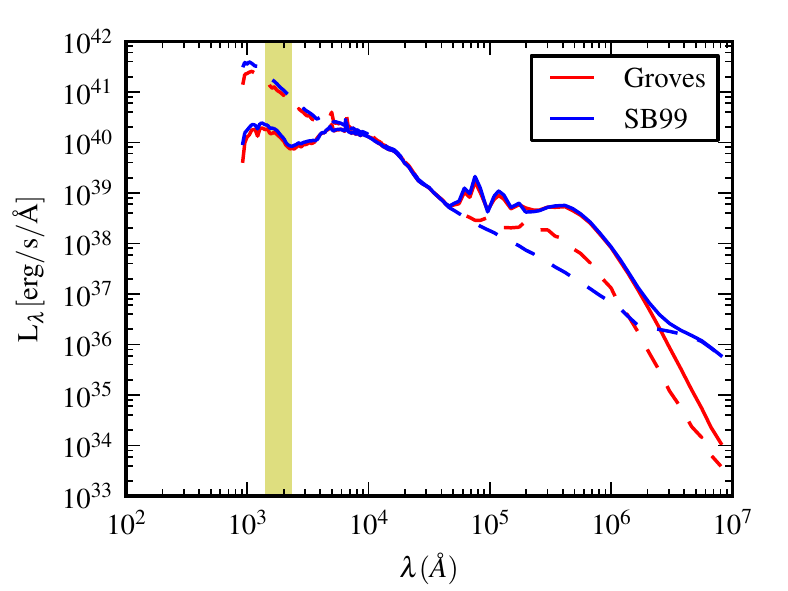}}
\caption{Sensitivity of our results to the use of the H\,{\sc ii} region model of \citet{Groves:2008ey}. In both panels,
the red (blue) lines show the results when the effect of H\,{\sc ii} regions is (not) included in the radiative transfer calculations;
all other assumptions are identical. The \emph{left} panel shows the \irxbeta\, evolution (note that the $\beta$ range shown
is much smaller than that shown in the \irxbeta\, plots in the main text). Including the effects of H\,{\sc ii} regions results in slightly
redder UV slopes, but the difference in $\beta$ is small ($\la 0.2$). IRX is essentially unaffected, except for very early times.
The \emph{right} panel compares the \rev{rest-frame} SEDs of the two runs at $t = 2.4$ Gyr. The intrinsic (attenuated) SEDs are indicated by the dashed
(solid) lines. The yellow region shows the wavelength range that is used to compute the UV slope.
The slight reddening due to H\,{\sc ii} regions is evident, and it is clear that $L_{\rm IR}$ is unaffected.
(N.B. The spurious upturn in the SB99 SEDs at $\sim 200~\mu$m is the result of a known bug in the version of {\sc Starburst99}
used only for this test, and it has a negligible effect on $L_{\rm IR}$.) This test shows that our conclusions are insensitive
to whether the \citet{Groves:2008ey} H\,{\sc ii} region model is employed.}
\label{fig:sps_model}
\end{figure}

The hydrodynamical simulations do not resolve the structure of the ISM on scales $\la 300$ pc; this represents one of the primary
uncertainties in the radiative transfer calculations (see, e.g., \citealt{H11}, \citealt{Snyder:2013}, L14, and \citealt{HS15}
for detailed discussions). To address this limitation, \sunrise treats sub-resolution dust clumpiness using two sub-resolution
models. First, to account for attenuation of young stars by their `birthclouds', the SED templates of
\citet{Groves:2008ey} are used when assigning input SEDs to star particles with age $<10$ Myr. These templates account for
dust attenuation, photoionization, and dust reemission in H\,{\sc ii} regions and photodissociation regions (PDRs). The time-averaged
PDR covering fraction, $f_{\rm PDR}$, is a key parameter that affects the resulting SEDs (see \citealt{Groves:2008ey} for details).
For our purposes, we note that higher $f_{\rm PDR}$ values result in higher attenuation and thus more dust reemission.
The second model accounts for sub-resolution dust clumpiness outside a star particle's `birthcloud' and is discussed in detail
in Appendix \ref{S:multiphase}. 

Both the $z \sim 0$ and $z \sim 2-3$ simulations use the \citet{Groves:2008ey} SEDs, but the $f_{\rm PDR}$ values differ: in the $z \sim 0$ simulations,
the default value, $f_{\rm PDR} = 0.2$ \citep{Jonsson10}, is used, whereas in the $z \sim 2-3$ simulations, $f_{\rm PDR} = 0$ is assumed.
\rev{Both sets of simulations use the fiducial \citet{Jonsson10} value for the cluster mass, $M_{\rm cl} = 10^5$ M$_{\odot}$, the other free
parameter in the \citet{Groves:2008ey} model. (This parameter has a negligible effect on the results compared with the
uncertainties investigated here; see fig. 18 of \citealt{Jonsson10}.)}
The reasons for these choices are discussed in detail in L14 and H13. It is possible that our results are sensitive to both use
of the \citet{Groves:2008ey} SEDs (rather than pure {\sc Starburst99}, hereafter SB99, SEDs) and the specific value of $f_{\rm PDR}$ employed.
To investigate the former possibility, we re-ran the radiative transfer calculations for the M3M3e merger simulation using SB99
SEDs for all star particles (rather than the default of \citealt{Groves:2008ey} SEDs for $<10$ Myr-old star particles and SB99 SEDs for
older star particles). We also re-ran the M3M3e calculations using the \citet{Groves:2008ey} SEDs for young star particles but assuming
$f_{\rm PDR} = 0$ rather than the default value for the $z \sim 0$ simulations, $f_{\rm PDR} = 0.2$. For both of these runs, the $z \sim 0$ default sub-resolution
treatment, `multiphase on' (see the next section), was used. The results of these two runs are compared in Figure \ref{fig:sps_model}.
The differences between the two are solely due to the effects of the \citet{Groves:2008ey} H\,{\sc ii} regions because PDRs are not included
and the \citet{Groves:2008ey} models use SB99 SEDs as input SEDs.

The left panel of Figure \ref{fig:sps_model} compares the evolution of the two runs in the \irxbeta\, plane. Again, the region
shown is a subset of the region shown in the \irxbeta\, plots in the main text because we wish to highlight the differences
between the runs. As expected, including the effects of H\,{\sc ii} regions (the red arrows) results in slightly redder UV slopes.
However, the difference in $\beta$ between the two runs at fixed time is at most $\sim 0.2$, which is small compared to
that associated with the assumed dust model and comparable to the variation with viewing angle. IRX is essentially unaffected,
except for very early times (when IRX is low), and even then, the difference is insignificant for our conclusions.

The right panel of Figure \ref{fig:sps_model} compares the SEDs of the two runs at $t = 2.4$ Gyr. The dashed lines correspond
to the input SEDs, whereas the solid lines denote the SEDs that result from
the radiative transfer calculations (i.e., they include dust attenuation and reemission from host-galaxy dust). The modest difference
in the UV slopes of the input SEDs (the SED that results when the \citealt{Groves:2008ey} templates are assumed has a slightly redder UV slope)
results in a slight difference in the observed UV slopes. The resulting IR SEDs are almost identical except for the
spurious upturn in the SB99 SED at $\sim 200~\mu$m, which has a negligible influence on $L_{\rm IR}$.
This is due to a known bug in the version of SB99 employed for this test, and the results presented elsewhere
in the paper are unaffected because the \citet{Groves:2008ey} SEDs, which are used for the fiducial runs, do not exhibit this feature.
Overall, the this test demonstrates that our results would not change significantly if we had employed pure SB99
SEDs rather than the \citet{Groves:2008ey} SEDs for $<10$ Myr-old star particles.

\begin{figure}
\centering
\vskip -0.0cm
\resizebox{3.0in}{!}{\includegraphics[angle=0]{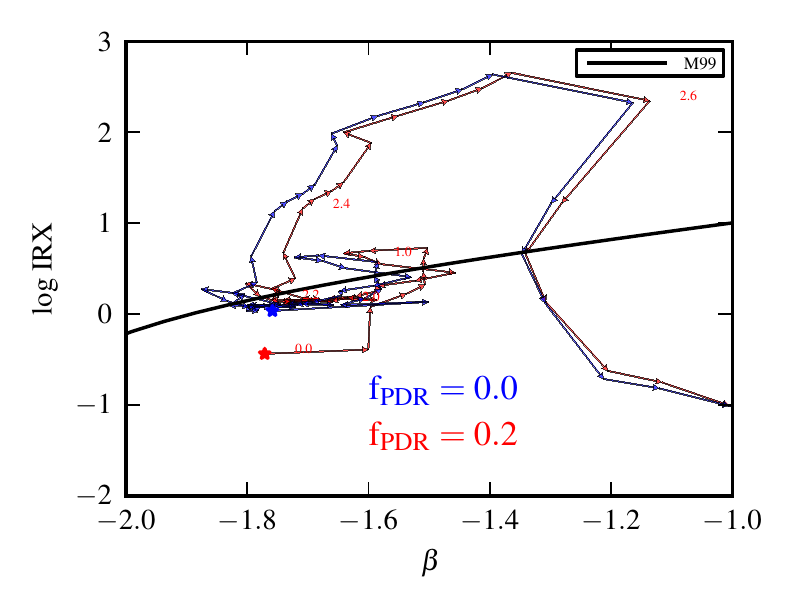}}
\caption{How the \irxbeta\, evolution of the M3M3e major merger depends on the assumed time-averaged PDR
covering factor, $f_{\rm PDR}$. The red (blue) arrows show the evolution when the fiducial value $f_{\rm PDR} = 0.2$
($f_{\rm PDR} = 0$, i.e., no PDRs) is used. All other parameters are identical. Including the effects of PDRs makes
the UV slope slightly redder, but the difference in $\beta$ is small ($\la 0.1$). IRX is unchanged except at early times,
and the difference then is modest. This test demonstrates that our results are insensitive to whether the effect of PDRs
is included in the input SEDs.}
\label{fig:fpdr}
\end{figure}

To investigate the effect of the sub-resolution PDR treatment, we compared the fiducial M3M3e run (\citealt{Groves:2008ey} SEDs for
young star particles, $f_{\rm PDR} = 0.2$, and `multiphase on' ISM treatment) with the aforementioned
run that used the \citet{Groves:2008ey} SEDs and the `multiphase on' treatment but assumed $f_{\rm PDR} = 0$. The differences
between these runs are thus solely due to the attenuation and dust reemission owing to the PDR model. Figure \ref{fig:fpdr}
compares the evolution of these two runs in the \irxbeta\, plane. Again, only a small range in $\beta$ is shown to highlight
the differences between the runs. At fixed time, inclusion of PDRs can make the UV slope slightly redder, as expected,
but the $\beta$ values differ negligibly (by $< 0.1$). IRX is noticeably affected only at very early times. This test demonstrates
that our results are insensitive to the assumed value of $f_{\rm PDR}$ because in the simulations, the attenuation from
\revtwo{`resolved' dust (i.e., that directly treated in the radiative transfer calculations)
tends to dominate over the attenuation inherent in the \citet{Groves:2008ey} sub-resolution PDR model} (fig. 9 of \citealt{Jonsson10}),
\revtwo{partially because a significant fraction of the UV luminosity can originate from stars older than 10 Myr, which are
attenuated \emph{only} by diffuse dust. However, it should be noted that} assuming an extreme value of $f_{\rm PDR} = 1$
(which we do not advocate; see \citealt{H11}) can significantly affect the UV slope (fig. 19 of \citealt{Jonsson10}).
Moreover, it is possible that the assumptions of the \citet{Groves:2008ey} model regarding e.g. the birthcloud
dispersal timescale do not hold. If so, simply varying the parameters of the model is insufficient to fully explore the
uncertainty in the amount of attenuation of UV light from dust on sub-resolution scales. Ultimately, fully addressing this issue
will require performing a similar analysis on ultra-high-resolution hydrodynamical simulations that resolve individual star-forming clouds
sufficiently well to directly capture both `local' and `diffuse' dust attenuation of UV light.

\section{Effects of assumption regarding the structure of the ISM on sub-resolution scales} \label{S:multiphase}

\begin{figure}
\centering
\vskip -0.0cm
\resizebox{3.0in}{!}{\includegraphics[angle=0]{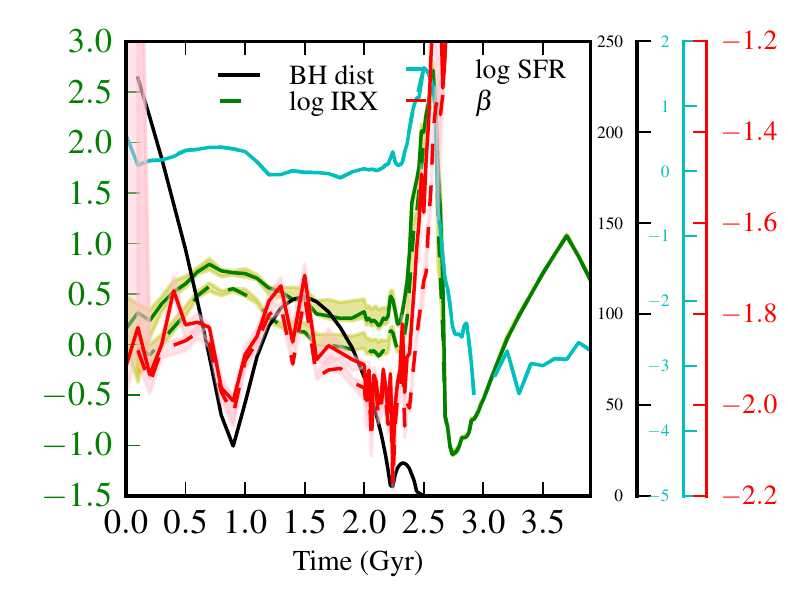}}
\resizebox{3.0in}{!}{\includegraphics[angle=0]{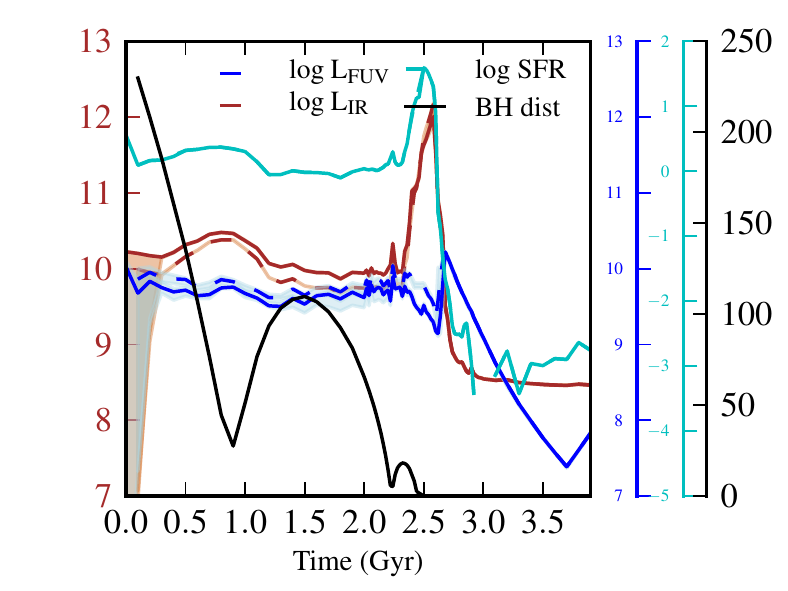}}
\caption{The effects of the treatment of sub-resolution dust clumpiness.
\emph{Left}: The time evolution of IRX and the observed $\beta$ for the default `multiphase on' (dashed lines) and `multiphase off' 
(solid lines) sub-resolution ISM models. The SFR and BH separation are also shown for reference. \emph{Right}: $L_{\rm IR}$,
$L_{\rm FUV}$, SFR, and BH separation versus time; the line styles are the same as in the left panel.
Pre-coalescence, using the `multiphase off' assumption results in slightly higher $\beta$, IRX, and $L_{\rm IR}$ and slightly
lower $L_{\rm UV}$, as expected (see text), but the differences are minor. Our conclusions are thus insensitive to the
treatment of sub-resolution dust clumpiness that we employ.}
\label{fig:ism}
\end{figure}

The second sub-resolution model address unresolved dust clumpiness outside of `birthclouds' because
the attenuation experienced by a photon packet traversing a given cell can be affected by unresolved clumpiness of the dust within the cell.
To characterize the uncertainty associated with this unresolved structure, \sunrise uses two extreme assumptions. In the first,
which has been referred to as `multiphase on' or `default ISM' in previous works, it is assumed that the cold clouds implicit in the
\citet{Springel03} model have zero volume filling factor. Thus, the dust content of these clouds is ignored in the radiative transfer
calculations. The alternate treatment, which has been referred to as `multiphase off' or `alternate ISM', assumes that the dust density
is constant within a given cell. The optical depth of a line of sight through a cell is minimized (maximized) under the multiphase-on
(multiphase-off) assumption \citep[e.g.,][]{1996aWitt}.

The default $z \sim 0$ ($z \sim 2-3$) simulations use the `multiphase on' (`multiphase off') ISM assumption; see L14 (\citealt{H11}
and H13) for justification of these choices. It is possible that our results are sensitive to this assumption. For this reason,
we re-ran the M3M3e radiative transfer calculations using the `multiphase off' assumption and keeping all other parameters
the same as in the fiducial run. The results are compared with the fiducial M3M3e results in Figure \ref{fig:ism}. The left
panel compares the time evolution of the observed $\beta$ and IRX values (and also shows the SFR and central supermassive
BH separation for reference), and the right panel compares the time evolution of $L_{\rm FUV}$ and $L_{\rm IR}$. In both
panels, the solid (dashed) lines show the results for the `multiphase off' (fiducial `multiphase on') run. Pre-coalescence,
using the `multiphase off' assumption results in slightly higher $\beta$, IRX, and $L_{\rm IR}$ values and slightly
lower $L_{\rm UV}$ values, as expected given that the `multiphase off' assumption maximizes the optical depth through
a given cell. However, the differences are very small, and our conclusions are thus insensitive to this assumption.

Note that in the right panel of Figure \ref{fig:ism}, during the starburst induced at merger coalescence, $L_{\rm IR}$ increases
sharply because of the elevated SFR, but $L_{\rm FUV}$ \emph{decreases} because of the increased attenuation
experienced by the centrally concentrated starburst. This illustrates why starbursts tend to cause the mergers to move
almost vertically in the \irxbeta\, plane and highlights the difficulty of inferring the SFR of highly dust-obscured systems
from the UV luminosity: if such systems are contained in a given UV-selected sample, applying a fixed dust correction
or using the M99 \irxbeta\, relation will cause the SFR to be significantly underestimated during the starburst. If
a significant fraction of the integrated stellar mass is formed in the burst, this underestimate can cause the inferred
SFR density to differ significantly from the true value.

\end{document}